\documentclass[12pt,preprint]{aastex}
\usepackage{epsfig}
\usepackage{graphicx}
\usepackage{epstopdf}
\newcommand   {\about} {\mbox{$\sim$}}

\renewcommand {\deg}   {\mbox{$^\circ$}}

\newcommand   {\kms}   {\mbox{km\,s$^{-1}$}}
\renewcommand {\ga}    {\mbox{\rlap{\hbox{\lower5pt\hbox{$\sim$}}}\hbox{$>$}}}
\renewcommand {\la}    {\mbox{\rlap{\hbox{\lower5pt\hbox{$\sim$}}}\hbox{$<$}}}

\begin{document}

%\title{ }
%\author{ }
%\email{ }
%\altaffiltext{1}{ }

%\begin{abstract}
%\vspace {5pt}
%\end{abstract}
%\keywords{ }
%\documentclass[12pt,preprint]{aastex}
%% manuscript produces a one-column, double-spaced document:
%% \documentclass[manuscript]{aastex}
%% preprint2 produces a double-column, single-spaced document:
%\documentclass[preprint2]{aastex}
%% Sometimes a paper's abstract is too long to fit on the
%% title page in preprint2 mode. When that is the case,
%% use the longabstract style option.
\def\kms {\hbox{km{\hskip0.1em}s$^{-1}$}} % km/s
\def\msol{\hbox{$\hbox{M}_\odot$}}
\def\lsol{\hbox{$\hbox{L}_\odot$}}
\def\kms{km s$^{-1}$}
\def\Blos{B$_{\rm los}$}
\def\etal   {{\it et al. }}                     % et al
\def\psec           {$.\negthinspace^{s}$}
\def\pasec          {$.\negthinspace^{\prime\prime}$}
\def\pdeg           {$.\kern-.25em ^{^\circ}$}
\def\degree{\ifmmode{^\circ} \else{$^\circ$}\fi}
\def\ee #1 {\times 10^{#1}}          % \ee p       10^p
\def\ut #1 #2 { \, \textrm{#1}^{#2}} % \ut unit p  unit^p   
\def\u #1 { \, \textrm{#1}}          % \u unit     unit
\def\nH {n_\mathrm{H}}

\def\ddeg   {\hbox{$.\!\!^\circ$}}              % Degrees over dot
\def\deg    {$^{\circ}$}                        % Degrees symbol
\def\le     {$\leq$}                            % <=
\def\sec    {$^{\rm s}$}                        % Second of time
\def\msol   {\hbox{$M_\odot$}}                  % Solar mass
\def\i      {\hbox{\it I}}                      % italic I
\def\v      {\hbox{\it V}}                      % italic V 
\def\dasec  {\hbox{$.\!\!^{\prime\prime}$}}     % Arcseconds over dot
\def\asec   {$^{\prime\prime}$}                 % Arcseconds symbol
\def\dasec  {\hbox{$.\!\!^{\prime\prime}$}}     % Arcseconds over dot
\def\dsec   {\hbox{$.\!\!^{\rm s}$}}            % Second over dot
\def\min    {$^{\rm m}$}                        % Minutes of time
\def\hour   {$^{\rm h}$}                        % Hours of time
\def\amin   {$^{\prime}$}                       % Arcminutes symbol
\def\lsol{\, \hbox{$\hbox{L}_\odot$}}
\def\sec    {$^{\rm s}$}                        % Second of time
\def\etal   {{\it et al. }}                     % et al.

\def\xbar   {\hbox{$\overline{\rm x}$}}         % bar over x

%\slugcomment{draft \#3}
\shorttitle{chemistry}
\shortauthors{zadeh}

\title{74 MHz Nonthermal Emission from Molecular Clouds: Evidence for 
a Cosmic Ray Dominated Region at the Galactic Center}

\author{
F. Yusef-Zadeh$^{1*}$,  M. Wardle$^2$, D. Lis$^3$, S. Viti$^4$, C. Brogan$^5$, 
E. Chambers$^6$, M. Pound$^7$ \&  M.~Rickert$^1$
} 
\affil{$^1$Department of Physics and Astronomy and Center for Interdisciplinary Research in Astronomy,
Northwestern University, Evanston, IL 60208, USA}
\affil{$^2$Department of Physics \& Astronomy, and Research Centre for Astronomy, 
Astrophysics \& Astrophotonics, Macquarie University, Sydney NSW 2109, Australia}
\affil{$^3$California Institute of Technology, MC 320-47, Pasadena, CA 91125, USA}
\affil{$^4$Department of Physics and Astronomy, University College London, Gower St. London, WCIE 6BT, UK}
\affil{$^5$National Radio Astronomy Observatory,  Charlottesville, VA 22903, USA}
\affil{$^6$University of Cologne, Cologne, Physikalisches Institut, Universit\"at zu K\"oln, 50397, Germany}
\affil{$^7$University of Maryland, Department of Astronomy, College Park,  MD 20742, USA}

\doublespace
\begin{abstract} 

We present 74 MHz radio continuum observations of the Galactic center region. These 
measurements show nonthermal radio emission arising from molecular clouds that is unaffected 
by free-free absorption along the line of sight. We focus on one cloud, G0.13--0.13, 
representative of the population of molecular clouds that are spatially correlated with 
steep spectrum ($\alpha^{74\rm MHz}_{\rm 327MHz}=1.3\pm0.3$) nonthermal emission from the 
Galactic center region. This cloud lies adjacent to the nonthermal radio filaments of the Arc 
near l$\sim0.2^{\circ}$ and is a strong source of 74 MHz continuum, SiO (2-1) and FeI K$\alpha$ 
6.4 keV line emission. This three-way correlation provides the most compelling evidence yet 
that relativistic electrons, here traced by 74 MHz emission, are physically associated with 
the G0.13--0.13 molecular cloud and that low energy cosmic ray electrons are responsible for 
the FeI K$\alpha$ line emission. The high cosmic ray ionization rate $\sim10^{-13}$ s$^{-1}$ 
H$^{-1}$ is responsible for heating the molecular gas to high temperatures and allows the 
disturbed gas to maintain a high velocity dispersion.  LVG modeling of multi-transition SiO 
observations of this cloud implies  H$_2$ densities $\sim 10^{4-5}$ cm$^{-3}$ and 
high temperatures. The lower limit to the temperature of G0.13-0.13 is  
$\sim100$K, whereas the  upper limit is  as high as 1000K.  
Lastly, we used  a time-dependent chemical model in which cosmic rays drive the 
chemistry of the gas to investigate  for molecular line diagnostics of cosmic ray heating. 
When the cloud reaches chemical equilibrium, the abundance ratios of HCN/HNC and 
N$_2$H$^+$/HCO$^{+}$ are consistent with measured values. In addition, 
significant abundance of 
SiO is predicted in the cosmic ray dominated region of the Galactic center. We discuss 
different possibilities to account for the origin of widespread SiO emission detected from 
Galactic center molecular clouds.

\end{abstract}

\section{Introduction}

The interstellar medium (ISM) of the nucleus of our Galaxy has unique 
characteristics. Gas clouds are subject to the strong tidal field of the 
nucleus and are required to have higher density than in the Galactic disk in 
order to become gravitationally unstable and form stars.  This region of 
the Galaxy hosts two components in its central few hundred parsec. One 
is the reservoir of molecular gas showing enhanced molecular emission 
with higher velocity dispersion ($\sim20-30$) \kms\, and gas temperature 
(50-200K) compared to elsewhere in the rest of the Galaxy$^{1-7}$.
The other is the 
prevalence of a mixture of thermal emission 
from ionized gas at a temperature of few thousand degrees and 
 nonthermal radio synchrotrone  emission 
from relativistic electrons
(e.g.; Nord et al. 2004; Law et al. 
2008). The synchrotron  emission is best viewed 
at low radio frequencies ($< 1$GHz) whereas thermal free-free emission 
is better detected at high frequencies ($> 1$GHz). Large-scale surveys 
of this region suggest that relativistic electrons and molecular gas
 co-exist and possibly interact with each other. 
However, the apparent 
correlation could result from chance coincidence along the 
 long line of sight towards  the 
Galactic center. 
We present the first evidence for 
low frequency nonthermal emission closely tracing molecular gas, 
establishing that cosmic ray electrons are 
physically associated with individual molecular clouds. 
We study one molecular  cloud G0.13--0.13 in detail and postpone the discussion 
of other Galactic center clouds to elsewhere.

The interaction between  cosmic-ray particles and molecular gas has several far reaching 
consequences. For one, cosmic rays play an important role in star formation processes as they
are the primary source of ionization in molecular clouds that are 
self-shielded from UV radiation field. The interaction of cosmic ray 
electrons heats the gas to a higher temperature, which increases the 
Jeans mass and causes the  initial  mass function (IMF) to become 
top-heavy$^{8-10}$. The higher
ionization fraction due to the impact of
these electrons  reduces  the damping of MHD waves and 
helps to  maintains  the  high
velocity dispersion of molecular gas in the nuclear disk.
In addition, this interaction 
strengthens the coupling of gas to the magnetic field  slowing  star formation
by  increasing 
the time scale for ambipolar diffusion  before the onset of gravitational 
collapse.  

Recent measurements indicate a vast amount of H$_3^+$ and 
H$_3$O$^+$ $^{11-14}$
 as well as high temperature molecular gas 
distributed in the Galactic center. 
The inferred 
minimum cosmic ray ionization rate,  $\zeta\sim 10^{-15}$\, s$^{-1}$\, H$^{-1}$, 
is one  to two orders of magnitude higher in the Galactic center 
region than in the Galactic disk$^{10,13}$. The elevated cosmic ray ionization 
rate will increase the ionization fraction of electrons in molecular 
gas, and drive ion-neutral chemistry. 
Another consequence is enhanced fluorescent FeI K$\alpha$ 
6.4 keV emission$^{15,16}$. This 
nonthermal X-ray emission is uniquely detected from Galactic center 
clouds and results from filling the inner K-shell vacancies of neutral 
iron created by the impact of low energy cosmic ray electrons or 
by irradiation by X-ray photons$^{17-20}$. 
A widely accepted model argues that the 6.4 keV line is
 a result of  irradiation by a
hypothetical transient source associated with Sgr~A*, which was active 
about 400 and again 100 years ago, and that we are seeing
this transient's light echo in the 6.4 keV emission$^{21}$. 
 We discuss another  scenario in which low energy cosmic rays
can contribute significantly in production of  the  FeI K$\alpha$  line emission from 
Galactic center clouds.

The Galactic center population of molecular clouds have physical 
conditions similar to hot cores, but with the exception of Sgr B2, there is no 
strong evidence for ongoing massive star formation 
throughout the inner few hundred parsecs of the Galaxy, a region  known as the 
central molecular zone (or CMZ)$^{22}$. 
Line emission from a variety of molecular species has been detected throughout the 
CMZ including those that are produced in the gas phase 
(e.g.; HCN, HCO$^+$, HNC, N$_2$H$^+$) as well as those processed on grain 
surfaces, e.g. SiO $^{6,7}$. 
 With the exception of the strong radiation field near Sgr A* and the Arches 
cluster, PDRs can not be very important in the dense self-shielded 
clouds. In addition, the X-ray flux from the Galactic center is too weak 
to qualify for the application of XDR models. Global heating by cosmic rays 
can explain why the 
gas temperature 
is significantly higher 
than 
the dust 
temperature in a large fraction of the gas and dust clouds in the nuclear 
disk. In the cosmic-ray dominated region of the Galactic center,  
cosmic-ray heating  should be more significant than ambipolar and turbulent 
heating$^{23}$. 
Thus, it is natural to examine  the 
chemistry of the gas in the cosmic-ray dominated region and attempt to identify 
molecular line diagnostics of cosmic ray heating.

%The origin of this fluorescent line can also be explained in the 
%context of 
%the irradiation of molecular clouds by a burst of X-ray flux from the massive black hole Sgr A* at the dynamical 
%center of the Galaxy.
%To establish the interaction picture of cosmic rays and molecular clouds and to determine accurately the 
%processes that are involved to explain the properties of molecular clouds in the center of the Galaxy, 

To this end, we present 
high resolution 74 MHz observations of the Galactic center and show the spatial correlation between  molecular 
gas and 100 MeV cosmic ray electrons. 
Enhanced 74 MHz emission 
appears to coincide   with a large concentration 
of molecular gas in the inner 5$^{\circ}$ of the Galactic center$^{24}$. 
Here, we focus on G0.13--0.13,  a representative molecular cloud,  
 and compare its 74 MHz emission with that of excited rotational 
transitions of SiO and CS molecules. This molecular cloud lies along the 
nonthermal filaments of the radio Arc near l$\sim0.2^{\circ}$, the most 
prominent network of magnetized filaments in the Galactic center. 
Molecular line emission from CO, CS, SiO, H$^{13}$CO$^+$ and CS suggests 
that this cloud has a high column density and gas temperature$^{25-27}$. 
The kinematics of CS line emission from G0.13--0.13 
suggests an expansion of molecular gas into the nonthermal filaments$^{25}$. 
CO observations
imply  gas temperature T$\ge70$K and column density 
N(H$_2$)=6--7$\times10^{23}$ cm$^{-2}$. Multiple transitions of NH$_3$ 
line emission from this cloud  have 
measured two temperature components giving a range T$\sim$25K and 
T$\sim125-200$K$^{28}$. The low temperature 
component of molecular gas is similar to the measured  
18--22\,K dust temperature in clouds in the inner 2$^{\circ}\times1^{\circ}$ of 
the Galaxy$^{29,30}$.

The structure of this paper is as follows. We first show  
74 MHz emission from Galactic center molecular cloud G0.13--0.13 
and estimate the variation of spectral index of 
nonthermal emission between 327 and 74 MHz. 
We then show the association of the nonthermal 
radio emission from G0.13--0.13 with FeI K$\alpha$ line emission at 6.4 
keV and estimate the cosmic ray ionization rate needed to produce the FeI 
K$\alpha$ line emission. We will also present new molecular line 
observations of four rotational transitions of SiO to determine 
the temperature and density of molecular gas in G0.13--0.13 more accurately. 
We consider the interaction of relativistic electrons with the gas and 
compute  the cosmic ray ionization rate, implied by the 74 MHz emission, 
 as a function of assumed magnetic 
field strength.
 We  present the dependence  of the total cooling rate of the 
gas on gas temperature for high molecular gas densities. Lastly, we study the chemical consequences 
of the interaction of cosmic ray electrons with the molecular gas by 
modeling the abundance ratios of several molecular species as a function 
of time. A time-dependent gas-grain chemical model$^{31}$ is
used to explore how abundance ratios of five 
representative molecular species vary with gas density and cosmic-ray ionization 
rate.

Our  study of G0.13--0.13 is not only relevant to 
understanding the  molecular component of the nuclear disk  of our Galaxy$^{8}$ 
but also to external 
galaxies and ultraluminous infrared galaxies (ULIRGs) where cosmic rays are thought to be 
the driving mechanism for star formation$^{9}$.

%These  
%relationship supports that picture that low energy comic ray electrons can produce FeI 
%K$\alpha$ line emission at 6.4 keV and provide independent measure of gas temperature for a given density.  

%Among other things, this correlation 
%provided a compelling evidence that enhanced fluorescent FeI 
%K$\alpha$ 6.4 keV emission from molecular clouds are explained by the bombardment of cosmic ray electrons.

\section{Observations and Data Reduction}

We imaged the Galactic center at  74 MHz and mapped molecular line 
emission  from several molecular clouds in
the Galactic center using Mopra, CSO and CARMA.  
We only present observations of G0.13--0.13 here and will describe 
details of observations of other Galactic center clouds elsewhere. 

\subsection{VLA}

The 74 MHz  radio continuum 
observations of the Galactic center were   described previously$^{24}$. 
Briefly, these measurements were taken in multiple configurations of the 
Very Large Array (VLA)
of the National Radio Astronomy Observatory\footnote{The
National Radio Astronomy Observatory is a facility of the National Science Foundation, operated
under a cooperative agreement by Associated Universities, Inc.} (NRAO),
  with a resolution of 114$''\times60''$ (PA$=-5^{\circ}$) 
and rms  sensitivity of $\sim 0.12$ Jy. 
These observations are unprecedented in spatial resolution and sensitivity  at these  low frequencies. 
Given the strong  frequency dependence ($\nu^{-2.1}$)  of  free-free absorption, 
foreground and embedded thermal sources  are seen in absorption against the strong
nonthermal emission from the Galactic center. The suppression of thermal emission due to high opacity of 
ionized gas  at 74 MHz  allows 
us to readily identify  nonthermal sources that emit at this frequency.  

\subsection{CSO}

    Observations of the SiO~(5--4) and (6--5) rotational transitions at 217.1 and 260.5~GHz, respectively, were 
carried out in 2010 September using the wideband 230 GHz ``Z-Rx'' receiver of the Caltech Submillimeter 
Observatory (CSO) on Mauna Kea, Hawaii. The weather conditions were good, characterized by a 225~GHz zenith 
opacity of $\sim 0.05-0.07$, equivalent to 1--1.5~mm of precipitable water vapor. Typical single sideband 
system temperatures at the relatively low elevation of the Galactic center were \about 250~K. The CSO FWHM 
beam size at the two frequencies is $\sim 34^{\prime\prime}$ and $28^{\prime\prime}$, respectively, and the 
main beam efficiency, as determined from total power observations of Jupiter, was $\sim 70\%$. Spectra were 
taken in the ``on-the-fly mapping'' mode, on a $\sim 15^{\prime\prime}$ grid, using a designated off 
position at 
$(\alpha, \delta)_{2000.0} =$\mbox{(17$^h$45$^m$13\psec 2,      
$-$28$\degree$45$^\prime$31\pasec 0)}, 
out of the Galactic plane. Multiple 
maps scanned in orthogonal directions were averaged together to avoid striping. 
The final per pixel 
integration time (ON) was 14--17 sec for SiO 5--4 and 25--31 sec for SiO 6--5. 
We used the high-resolution 
CSO facility FFTS spectrometer with 8192 channels and a total bandwidth of 1~GHz.

\subsection{Mopra}
We observed G0.13--0.13 on 2010,  June 28  
using Mopra\footnote{
The data was obtained using the Mopra radio telescope, a part 
of the Australia Telescope National Facility which is funded 
by the Commonwealth of Australia for operation as a National Facility managed by CSIRO. The University of 
New South Wales (UNSW) digital filter bank (the UNSW-MOPS) used for the observations with Mopra was 
provided with support from the 
Australian Research Council (ARC), UNSW, Sydney and Monash Universities, as well as the CSIRO.} in
the on-the-fly mapping mode.  The map size is 5$'$\, by 5$'$, and is
centered at Galactic longitude and latitude (l, b)=($0.1092^{\circ}, -0.1000^{\circ}$).  We used a latitude
scan direction (orthogonal to the Galactic plane), and a row spacing
of $\sim15''$.  We used the MOPS spectrometer in Wideband mode,
with a central frequency of 89.690~GHz.  This results in a frequency
coverage of $\sim~85.6 - 93.8$GHz  across four sub-bands.  Each
sub-band has 8192 channels, a channel width of 269.5~kHz
($\sim0.9$\,   km~s$^{-1}$), and two linear polarizations.  The final map
is the average of the two polarizations. 
Regular pointing checks done throughout the
observations, and were within $\sim$~4$''$ (the beam size is $\sim$~38$''$\, at these frequencies). 
 The intensity is given 
in T$^*_A$ which can be  divided  
by the main beam efficiency  0.49 to convert the intensity to T$_{mb}$. 

We also made deep pointed observations toward four positions (A-D) within G0.13--0.13
on  2010, June 30 (see Table 1).
These spectra were
obtained in an on-off position switching mode 
for a total on-source time of 288~s.  The MOPS spectrometer was configured in wideband mode,
identically to the setup of the map described above.  
A total of 16 molecular lines CH$_3$CN, C$^{13}$CN, $^{13}$CS,
N$_2$H$^+$, HNC, HCCN, HNCO, HCN, HCO$^+$, $^{28}$SiO, HC$^{15}$N, SO, H$^{13}$CN, 
H$^{13}$CO, SiO, NH$^{15}$C and C$_2$H are detected from four positions in G0.13--0.13 (see Fig. 7).

%(see Table~\ref{point_obs}).  

\subsection{BIMA}

% a 6.042 -4.752 
% c 6.79  -6.504
% d 7.794  -7.752

We mapped G0.13--0.13 in   CS(2-1) at
97.980968 GHz and  HCO$^+$(1-0) at 89.188518 GHz
 with the Berkeley-Illinois-Maryland Association (BIMA)
interferometer\footnote{The BIMA interferometer was operated under a
joint agreement between the University of California, Berkeley, the
University of Illinois, and the University of Maryland with support
from the National Science Foundation.} during the 2002-2003 observing
season.  The BIMA array was a mm-wavelength interferometer located in
Hat Creek, California$^{32}$, 
consisting of ten 6.1~m
antennas.  These antennas have since been combined with
the Owens Valley Radio Observatory 10~m antennas into the
CARMA\footnote{Combined Array for Research in Millimeter-wave Astronomy;
~http://www.mmarray.org}
millimeter array at Cedar Flat, California.

The cloud was observed with 20 mosaicked pointings on a hexagonal grid
corresponding to Nyquist sampling of the primary beam and
covering a total area of about 5\arcmin~$\times$~6\arcmin.  The pointing   
center was 
$(\alpha, \delta)_{2000.0} =$\mbox{(17$^h$46$^m$24\psec 0,      
$-$28$\degree$45$^\prime$00\pasec 0)} 
with a velocity of 30 \kms\/ with
respect to the local standard of rest.
The HCO$^+$ and CS data were obtained simultaneously in opposite
sidebands.
Each of the spectral lines were observed with a 50.0 MHz correlator
window with 256 channels. Along with the spectral data, continuum
windows with a total coverage of about 800 MHz were also observed.
We observed in the C and B configuration of the array, sampling spatial
frequencies from 1  to 78 k$\lambda$.  Absolute flux calibration
was derived from observations of Mars or Uranus immediately before or after
the source track.  The quasar 1733-130 was used as the phase calibrator
and secondary flux calibrator.
To fill the inner {\it uv} spacings, we also obtained fully-sampled maps
covering the region imaged with the array in the HCO$^+$, and CS  using
the Five Colleges Radio Astronomy Observatory 14~m telescope,  as well as the 
 previously published CS(2-1) maps
made with Nobeyama 45-m radio telescope (NRO)$^{25}$. 
A more detailed high resolution kinematic study of G0.13--0.13 including the 
results of HCO$^+$ observations
will be given elsewhere.

%, which has a
%58.3\arcsec\/ FWHM beam at 86 GHz. 
%without being contaminated by thermal emission. 

\section{Results}

Figure 1a,b shows the distribution of 74 MHz continuum and  the CO (3-2) line 
from the inner 1.5$^{\circ}\times12'$ 
of the Galactic 
center. A layer of molecular gas in the CMZ runs parallel to the Galactic plane 
and is seen 
in the distribution of CO (3-2) line emission$^{33}$.  
This layer includes  some of the most prominent molecular clouds in the CMZ 
such as 
Sgr B2 and Sgr B1, the 40, 50, --30 and 20 \kms\, Sgr A and Sgr C clouds. 
Some of the prominent clouds  are labeled on Figures 1a,b. 
At positive longitudes, the Sgr B2 complex, G0.36-0.09 and G0.13--0.13
have 74 MHz counterparts and at negative longitudes 
the layer of molecular gas that runs parallel to the Galactic plane G359.75-0.23 
shows a 74 MHz counterpart. 
Another presentation of 
the  CO (3-2) and 74 MHz emission is shown in Figure 1c. 
The distribution of 
74 MHz emission appears to follow the  CO (3-2) line emission  except in two types  of regions. 
One type  is associated with 
holes created by free-free absorption suppressing  background 74 MHz emission and the other is where 
nonthermal continuum sources  have no molecular counterparts.
Overall, the 74 MHz emission shows  similar morphology to 
that of CO (3-2) line emission apart from absorption  features 
manifesting as holes in the 74 MHz distribution.
To clarify the correlation 
between absorption  features and free-free emission,
Figure 1d  presents a continuum  map at 8.5 GHz 
based on GBT observations$^{34}$.
Prominent continuum sources are labeled on this figure.  
Holes  in the 74 MHz  continuum map 
coincide with strong continuum emission at 8.5 GHz, 
consistent with the free-free  absorption coefficient 
increasing  significantly at low frequencies.  Thus,  thermal sources become optically thick 
and are seen in absorption against strong nonthermal background emission$^{35}$
whereas nonthermal sources that are not contaminated by significant foreground thermal emission 
are clearly identified at 74 MHz.  
The morphological comparison of these figures 
suggest  that nonthermal continuum emission at 74 MHz 
arises from  Galactic center molecular clouds, implying that  
cosmic-ray electrons  are interacting with molecular gas in this unique region of the 
Galaxy.

%To examine the spatial 
%correlation of low frequency and molecular line emission in more detail, 
%Figure  2a,b shows the distribution   of  74 
%MHz and  24$\mu$m images at positive longitudes. 
%A layer of gas known as the dust ridge appears as a continuous 
%chain of  clouds  running between 
%G0.25-0.0 and Sgr B2. These clouds have strong submillimeter counterpart 
%tracing the dust distribution$^{14}$. 
 %(Lis and Carlstrom 1994).  
%This gas layer is seen as   infrared dark 
%clouds (IRDCs) at 24$\mu$m  
%stret$ching between the radio Arc near G0.2-0.0  and G0.6-0.0 where
%Sgr B2 is located.  These IRDCs  appear to coincide with 
%regions of  enhanced 74 MHz continuum emission with the exception of the region 
%where continuum  emission from Sgr B2/B1 is detected. 
% Remarkably, the molecular gas to the SE of the continuum source  Sgr B2 , drawn schematically as a
%balloon-shaped structure,  appears to be spatially correlated with the distribution of the 74 MHz emission. 
%The bright source G0.9+0.1, a pulsar wind nebula$^{36}$, %(Camilo et al. 2009),  
% and G0.38+0.06 are nonthermal continuum sources and have no molecular or 
%dust counterparts. 

We now focus on  the G0.13--0.13 cloud. 
Figure 2a shows contours of CS (1-0) molecular line emission from this cloud 
superimposed on a grayscale 5~GHz 
continuum image. 
A network of linearly polarized filaments of the radio Arc running 
almost perpendicular to the Galactic plane at a PA$\sim-170^{\circ}$ lies at the eastern 
edge of the cloud$^{37}$. 
The morphology of 
molecular line and radio continuum emission from G0.13--0.13 
suggests that the molecular gas is surrounded by, and is interacting with,  
nonthermal radio filaments.  The kinematics of the molecular gas suggests that expansion of 
the cloud is responsible for a dynamical interaction between the eastern limb of the G0.13--0.13 and 
the magnetized nonthermal filaments$^{25}$. 
 Figure 2b shows  high resolution 
CS (2-1) line emission from the G0.13--0.13 molecular cloud (red) superimposed on a 1.4 GHz 
continuum image (green).  The molecular gas distribution is clumpy with the morphology 
of a ``boot",  and is edge-brightened parallel to the nonthermal radio filaments of the 
Arc near l$\sim0.2^{\circ}$ on the eastern edge of the cloud. We note that the western edge 
of the cloud lies parallel to another magnetized nonthermal filament G0.087-0.087$^{38}$. 
 The molecular line images show a ridge of emission that deviated from a straight line perpendicular to the 
Galactic plane (labelled as ``Meanderning feature'' on Figure 2a) 
near l$\sim8'$, b$\sim-7'$. This ridge of emission  
 appears to curve around a nonthermal filament near l$=8'\, 4''$\,, b=$-6'\, 55''$. 
This high resolution map of molecular line emission provides the strongest morphological 
support for the interaction of the nonthermal filaments and the eastern edge of G0.13--0.13.

Figure 3a,b  show contours of 74 and 327 MHz emission, respectively, 
superimposed on a grayscale image at 5 GHz. 
The 74 MHz emission is 
produced mainly by nonthermal processes  whereas 5 GHz and 327  MHz emission 
can trace emission produced by  both thermal and nonthermal processes. 
The distribution of 74 MHz emission  differs  remarkably from the 327 MHz and 5 GHz continuum. 
The 74 MHz emission arises  from  the vertical filaments as well as
the region where  the G0.13--0.13 molecular cloud lies. 
There is no evidence of thermal emission from G0.13-0.13 at 327 MHz or at
higher frequencies, thus, the spectrum derived here
is not  contaminated by thermal emission.
The 
vertical filaments seen in the grayscale image at 5 GHz 
have counterparts at 74 MHz except in the thermal region to the north of the filaments near G0.18-0.04. 
This is because G0.18-0.04 is an HII region dominated by free-free 
emission at high frequencies, thus the lack of 74 MHz emission is due to significant free-free 
absorption$^{35}$.

%These figures  show 
%nonthermal low frequency radio emission coincident with a molecular cloud. 

What is remarkable about  images shown in Figure 3a,b 
is the discovery 
that enhanced nonthermal emission  
at 74 MHz arises from the  molecular cloud  G0.13--0.13. 
The spectrum of  emission from the center of the cloud 
 must be steep given that there 
is no strong 327 MHz emission from G0.13--0.13. 
Figure 3c shows the spectral index distribution 
(where the flux density  F$_{\nu} \propto \nu^{-\alpha}$) between 327 and 74 MHz. 
The spectrum is inverted ($\alpha < 0$) at the position of the filaments and is steep    ($\alpha\sim1.3$) 
where the molecular line and 74 MHz emission peak. 
There is no evidence of thermal emission from G0.13-0.13 at 327 MHz or at 
higher frequencies, thus, the spectrum derived here 
is not  contaminated by thermal emission.

Another version of the spectral index distribution is made by comparing 
the background subtracted profiles of the emission at 74 and 327 MHz. 
Figures 3d,e show two different cross cuts made at constant longitude 
l=$-6'$ and constant latitude b=$-8'$, respectively. This technique 
follows earlier work by Law et al.  (2008).  The spectral index 
distribution, as shown in Figure 3d, is consistent with a steep spectrum 
($\alpha=1.1\pm0.2$) becoming flatter or inverted at the nonthermal 
vertical filaments. Figure 3e shows $\alpha\approx-0.2\pm0.05$ between 
74 and 327 MHz at the position of nonthermal filaments.  The electron 
energy spectrum of the center of G0.13--0.13 is remarkably steep, 
p$\approx$3.2 (where p=$2\alpha$+1 corresponding to energy spectrum 
E$^{-p}$). In contrast, the spectral index of the vertical filaments is 
remarkably flat  p$\approx$0.6. 
The variation of the spectral index of 
nonthermal emission from the filaments and G0.13--0.13, is 
$\Delta\alpha\sim 1.5$.

G0.13--0.13 is one of the Galactic center molecular clouds showing bright 
FeI K$\alpha$ line emission at 6.4 keV$^{39}$. 
Figure 4  shows contours of 
fluorescent FeI K$\alpha$ line emission 
based on Chandra observations convolved to a resolution of 
30$''$ and 
superimposed on 
a grayscale distribution of  integrated CS (1-0) line emission. 
High resolution Chandra  and CARMA 
images of this cloud show a clumpy distribution of K$\alpha$ line 
emission$^{20}$ tracing CS (1-0) line emission. 
The overall 
distribution of FeI K$\alpha$ line emission arising from G0.13--0.13 is 
similar to that of low frequency radio emission molecular line CS and 
SiO emission.  The distribution of the  K$\alpha$ line emission is 
clumpy and shows a ridge of emission 
adjacent to a nonthermal feature 
at l=8$'\, 10''$\, b=$-6'\, 55''\,$. Remarkably, this X-ray  ridge
feature is similar to that of  CS emission  (cf. Meandering feature in Fig. 2b). 
Both the X-ray and CS ridges 
curve around a nonthermal filament. The morphology of 6.4 
keV line emission  supports the idea that nonthermal radio filaments are interacting 
with the edge brightened FeI K$\alpha$ emitting molecular cloud 
G0.13--0.13 at 6.4 keV. 
These images suggest a three-way correlation 
between cosmic ray electrons, molecular gas and FeI K$\alpha$ line 
emission.

%Both sets of contours are 
%superimposed on a grayscale 5 GHz continuum image. 

%We also 
%investigated the physical association of molecular line and nonthermal continuum emission by 
%correlating the distribution of 74 MHz flux and HCN (1-0) molecular line emission. Figure 2d shows 
%this correlation by the diagonal curves supporting the argument that nonthermal emission arises from 
%the G0.13-0.13 molecular cloud.
%It turns out that  nonthermal emission from G0.13-0.13 is not an 
%isolated molecular cloud. 
%Using 
%The temperature and density of G0.13-0.13 have  been estimated based on 
%SiO (1-0) and 2-1 as well as the line ratios of CO 2-1 and 1-0 line emission. 
%These estimates suggest that the gas temperature is relatively warm 
%having a high column density of 6$\times10^{23} \rm cm^{-2}$.

In order to estimate 
the gas density and temperature in  G0.13--0.13,  
we  measured  the line intensities of  four transitions of SiO (6-5),  (5-4), (2-1) and (1-0). 
We used the SiO (1-0) and (2-1) images taken from 
Handa et al. (2006) and convolved them to the same resolution and velocity coverage, 
between  -10  to 50 \kms\,   
as those of SiO (6-5) and (5-4) transitions. 
Figure 5a shows contours of  velocity integrated SiO (5-4) line emission 
mapped over  a limited area  of G0.13--0.13 superimposed on a 
5 GHz continuum image.  We applied a
 Large Velocity Gradient (LVG)
model to the two E and NW rectangular boxes, as shown schematically,  
in order to  derive  the physical parameters of molecular  gas in G0.13--0.13. 
The LVG grid was computed for a constant SiO column density of 1.6$\times10^{13}$ cm$^{-2}$
and a line width of 30 \kms. 
This approximately reproduces the observed line intensities for the range of densities and temperatures applicable to the
gas in the Galactic center region.
With these assumptions, the emission is optically thin (typical optical depths ranging between
0.1 and 10$^{-3}$, 
depending on the transition). Thus,  
optical depth effects do not influence the LVG model results as long as the beam filling factor 
is close to unity.  A smaller beam filling factor will change the optically thin assumption we have made. 
Figure 5b,c  show  the density and kinetic  temperature constraints  
estimated from three SiO line ratios  of  the E and NW 
boxes in three different colors, respectively. 
The three line ratios give a wider range of   gas temperature between 
$\approx100-1000$K
for  box E and  and $\approx400-1000$K for box NW. 
A  lower range of 
gas temperatures  $>100$K is found  from the region near the nonthermal filaments of the Arc,  box E as
shown in Figure 5a,  
than that  of  the   peak emission in the  NW box having T$>$400K. 
The implied  kinetic temperatures are consistent with earlier measurements based 
on multiple transitions of NH$_3$ $^{28}$. 
Additional constraints  have  also been made  by estimating 
  molecular gas pressure from  the NW box  of Figure 5a.  
 The top three  panels of  Figure 5d show the plots of 
 line ratios for all models  in the grid as function of logarithm of pressure  (nT). 
The bottom  panel shows a sum of 
[(model-observed)/observed]$^2$ for the three line ratios and the the best fit 
is obtained for log(nT) in the range of 6.4--6.9. The line ratios give values of 
density and temperature that are consistent with  log(n)$\sim4.4$ and log(T)$\sim2.3$.  
for minimum $\chi^2$ values. 
Similar values were obtained  for the E box  of Figure 5a.

Tsuboi et al. (2011) used the SiO (2-1)/(1-0) line ratios to estimate 
a molecular gas density of $\sim10^4$ cm$^{-3}$ at a kinetic temperature of 60K. 
On the other hand, Oka et al. (2001) used the CO (2-1)/(1-0) intensity ratios 
and argued that G0.13--0.13 has a low  density $\sim10^2$ cm$^{-3}$, 
high kinetic temperature $\ge70$K component. 
Overall, our  model fits which uses 
multiple transitions of SiO   are  suited to measure 
gas density and temperature, though 
with  systematic uncertainties from using  different  telescopes. 
The application of 
the LVG  excitation code constrains the gas density of hydrogen nuclei 
 n$_{\rm H}\sim (1-3)\times10^4$ cm$^{-3}$ 
and temperature T$\sim (1-2)\times100$K but we can not rule out gas temperature as much as
1000K and lower densities  n$\sim10^3$ cm$^{-3}$. 

Two studies have made discrepant estimates of
the column density of the gas in G0.13--0.13 $^{26,40}$
Using  kinetic temperature  T$_k$=70K,
Handa et al. (2006) used H$^{13}$CO$^+$ (J=1-0) line emission which
is expected to be optically thin with abundance of 10$^{-10}$ to derive an 
H$_2$ column density (6-7)$\times10^{23}$ cm$^{-3}$.
On the other hand, Amo-Baladron et al. (2009) determined the physical condition
 by applying
an  LVG model to the 
SiO (2-1) and (3-2)  transitions and  then  estimated  the column of H$^{13}$CO$^+$ (J=1-0).
Using the same abundance as that of Handa et al. (2006),
the derived  column density of molecular hydrogen from this study is
(2-8)$\times10^{22}$ cm$^{-3}$.
To resolve this discrepancy, we estimated the column
density of molecular gas by using submillimeter emission from dust grains in G0.13--0.13.
We used the data taken from
Pierce-Price et al. (2000) who derived 513 \msol\, Jy$^{-1}$ beam$^{-1}$ at 850$\mu$m with the assumption that dust
temperature is 20K, metallicity twice solar and opacity index $\beta$=2.  Using background subtracted flux densities of 0.5
and 1.5 Jy from the NW and E boxes at 850$\mu$m and a beam of 15$''$, we find the column density of molecular hydrogen
ranges between
$\sim3.5\times10^{22}$ and
$\sim10^{23}$, cm$^{-2}$ 
respectively.
There is  uncertainty in accurately determining 
 the background emission. Given this uncertainty, 
 our estimate of the column density falls between previous measurements of
Handa et al. (2006) and  Amo-Baladron et al. (2009).
Herschel   maps presented by Molinari et al. (2011) 
give column density values that are  consistent with our estimates. 

% (as best as we can determine from 
%their maps).

%Past molecular line surveys have  have considered a correlation  the line intensity ratio  SiO (2-1)/ CS (1-0)  
%and the 6. 7 keV line emission. 

% The gas 
%density and temperature  ranges 
% are within  $\sim10^{5-6}$ cm$^{-3}$ 
%and  150-200K, respectively. 
%T$\sim10^{2.15}$K and n$\sim10^{5.5}$ cm$^{-3}$.
% are consistent with previous estimates. 
%SiO (5-4)/(6-5)   and SiO (2-1)/(5-4) line ratios. 

%Figure 4a-c 
%show three images of a larger region
%surrounding the molecular complex 
%Sgr B at 24$\mu$m, 74 MHz and 6.4 keV emission, respectively. 
%The 24$\mu$m image shows prominent infrared dark clouds to the north of 
%thermal emission from the Sgr B complex, whereas 74 MHz shows 
%diffuse and faint emission from IRDCs. There are two other bright 
%nonthermal sources 
%that have no molecular counterparts. The source G0.9 is a known 
%supernova remnant.  The 6.4 keV line emission also follows IRDCs and 
%74 MHz emission  in the region where thermal emission is 
%not dominated. 

\section{Discussion}

Our interest in studying Galactic center molecular clouds stems from three earlier studies, all of which 
had suggested the interaction of cosmic ray electrons with molecular gas and that cosmic ray electrons 
are enhanced in 
this region. First, fluorescent 6.4 keV 
K$\alpha$ iron line emission from several molecular clouds in the Galactic center 
 may be  the consequence of the interaction of low energy cosmic ray electrons 
with molecular gas$^{20,41}$, 
or alternatively, 
the irradiation of molecular clouds by a hypothetical X-ray flash  associated with past activity of Sgr~A*\,$^{21}$.
Second, nonthermal bremsstrahlung 
from this electron population can also  explain the diffuse 
$\gamma$-ray emission from the central 2$^{\circ}\times1^{\circ}$ of the Galactic center$^{42}$. 
Third, the strong H$_3^+$ absorption along several lines of sight towards the Galactic 
center imply  that the  ionization rate 
is at least an order of magnitude higher than elsewhere in the Galaxy$^{10}$. 
This is consistent with the low energy cosmic ray electron production of 
the FeI K$\alpha$ line emission. 
Finally,  low frequency diffuse radio emission$^{43}$ 
combined with spectral index 
measurements$^{23}$ 
suggest strong nonthermal  continuum emission from the Galactic center.  
Our detailed study 
provides  the most
compelling evidence  yet that 
relativistic electrons, here  traced by 74 MHz emission,   are physically associated with 
the G0.13--0.13 molecular cloud. 
This  cloud  is  one of the strongest sources 
of FeI K$\alpha$ line emission in the Galactic center. 
We complete  our analysis by 
 studying the 
complex morphology of G0.13-0.13,  estimating  the cosmic ray ionization rate,  
the expected FeI K$\alpha$ line emission at 6.4 keV,  and modeling 
chemical signature of the interaction.

%A more detailed account of the application of the model to other Galactic center 
%molecular clouds  will be given elsewhere. 

\subsection{Morphology \& Kinematics}

G0.13--0.13 is a complex  molecular cloud  with kinematics that 
suggest  expansion  into the vertical filaments of the radio Arc$^{25}$ 
Morphologically, this cloud lies at the center of a circular-shaped 
structure, known as the 
radio  Arc bubble, apparent  in mid-IR images taken with the Spitzer and  MSX observatories$^{44-47}$. 
Figure 6 shows contours of SiO (2-1) line emission, obtained with Mopra,  
superimposed on a three-color Spitzer IRAC image. The Arc bubble is noted exterior to 
the south of G0.13--0.13. 
The origin of the bubble and its relation to the radio Arc 
and the G0.13--0.13 cloud  
is not well understood. Past studies have suggested 
that the radio Arc bubble lies in the vicinity of the  Quintuplet star cluster and 
  is  produced by stellar winds 
or supernova explosions sweeping up interstellar material$^{47}$. 
The nonthermal radio emission  at 74 MHz and 
expansion 
of G0.13--0.13 point to the possibility that 
the Arc bubble may be  produced by the same event.  
It is possible that 
an energetic event is  
driving a shock into G0.13--0.13.  The shock  reaches 
the edge of the cloud and before it encounters  much lower density of the gas
exterior to G0.13--0.13,  then  sweeps through the low density gas and creates a 
edge-brightened bubble. 
In this picture, the expansion of  G0.13--0.13 into the   nonthermal vertical filaments 
accelerates  particles along the filaments$^{25,48}$, 
thus generating 
a young population of electrons running along the magnetized filaments with an  unusually 
flat    energy spectrum (p$\sim$0.6). 
The puzzle, however, is  the origin of  the  steep 
  energy   spectrum of particles with p$>3$ in  G0.13--0.13. 
This is because  
energy losses  are  increasingly severe for lower energy electrons, 
tending  to   flatten their energy spectrum   at low energies. 

%One non-standard model that we mention is the possibility that 
% a new population of  cosmic  ray electrons and positrons 
%at low energies  are  created through the annihilation of 
%a relatively light  dark matter particle. 
%The annihalation  can provide a new population of electrons at
%energies less than the annihilation energy of WIMPS 
%(Linden, Hooper and Yusef-Zadeh 2011). In this picture, 
%all Galactic center clouds should show the same steep spectrum 
%of electrons as that noted toward G0.13--0.13. 

%A more detailed account of these suggestions 
%will be given elsewhere. 

%One possibility is that the magnetic field in G0.13--0.13 is strong enough that 
%high energy electrons lose their energies rapidly, thus the spectrum of the particles become 
%soft at low energies. 

%As shown earlier, nonthermal diffuse emission at 74 MHz 
%arises from G0.13--0.13 as well as from radio filaments of the Arc.  

\subsection{Cosmic Ray Ionization Rate}

Using  equation (3) of Yusef-Zadeh et al. (2012), 
the cosmic ray ionization per second per hydrogen nucleus, $\zeta$,
depends  on
the observed synchrotron intensity I$_{\nu}$, 
the magnetic field
B, the depth of the source of emission along the line of sight, 
L, and the energy  spectrum  of the electrons with the index p,
assumed to be power-law E$^{-p}$ between 0.1 MeV  and 1 GeV 
and $\alpha = (p-1)/2$ is the synchrotron spectral index 
and $I_{\nu}\propto\nu^{-\alpha}$. 

\begin{equation}
\zeta \approx \frac{3.1\times10^{-14}}{p-1}\,
\frac{I_\nu}{\u Jy \ut arcmin -2 }\, \left(\frac{\nu}{\u GHz }\right)^{\!\alpha} \, \left( \frac{L}{30\u pc
}\right)^{-1}\, \left(\frac{B}{100\,\mu\mathrm{G} }\right)^{\!\!-(1+\alpha)}\, \ut s -1 \ut H -1
\end{equation}

Figure 7a shows the inferred cosmic ray ionization rate as a function of the magnetic field
strength for different values of $\alpha$.   
We assumed that L$\sim$2.5 pc  (1$'$ corresponds to 2.4pc at the Galactic center distance 8.5 kpc)
and the    observed   surface brightness of 0.71    Jy arcmin$^{-2}$ at 74 MHz 
averaged  over  the inner $5'\times5'$ of  G0.13--0.13. 
The black dot on the curve gives the equipartition magnetic
field for the electron population producing 
diffuse synchrotron emission in the dense molecular cloud  G0.13--0.13. 
The ionization rate due to electrons increases for steeper 
spectral index values  for 
a fixed emissivity at the observed frequency, as there is a successively larger population of 
lower-energy electrons radiating at lower frequencies.  
The equipartition magnetic field  ranges between 30 $\mu$G and 0.3 mG for 
$\alpha$ values between 0.25 to 1.5.  The corresponding          
$\zeta$ is estimated to be $\sim4\times10^{-15}$ and 10$^{-11}$ s$^{-1}$, respectively.  
A  cosmic ray ionization rate $\ge10^{-13}$ s$^{-1}$ H$^{-1}$  is
  sufficient  to fully dissociate the  gas on a time scale of 
$\leq10^6$ years.  
To avoid this, the magnetic field 
has to be larger than 1mG for $\alpha=1.25$. 

%Future magnetic field Zeeman measurements of this cloud will be able to 
%constrain the value of the  cosmic ray ionization rate.  

\subsection{The 6.4 keV Neutral Iron Line Emission}

We apply the cosmic ray  model  to the G0.13--0.13 molecular cloud  
by    estimating the   K$\alpha$ line emission from   the
interaction of the low energy 
nonthermal electrons responsible for  synchrotron radio emission detected at 74 MHz.
Using equation (6) of Yusef-Zadeh et al. (2012),
\begin{equation}
I_{\mathrm{K}\alpha}  \approx 8.7\ee -8 \left(\frac{\zeta}{10^{-14}\ut s -1 }\right)
\left(\frac{N_\mathrm{H}}{10^{23}\ut cm -2 }\right) {\rm \ ph \rm \ s^{-1} \ cm^{-2} \ arcmin^{-2}}\,,
\end{equation}
for N(H$_2$)=$(1-3)\times10^{23}$ cm$^{-2}$, 
then the cosmic ray ionization rate  needed to give
the background subtracted K$\alpha$ line 
flux of  $3\times10^{-6} \u ph\, \rm  cm^{-2}\, \rm s^{-1}$\, arcmin$^{-2}$ 
derived from  Suzaku observations$^{39}$ 
is $\zeta\sim(1-3)\times10^{-13}$\, s$^{-1}$, respectively. 
A  value of cosmic ray ionization rate 
$\zeta\sim10^{-13}$\, s$^{-1}$ 
is indicated  by  a dashed line on Figure 7a. 
Using the spectral index values between $\alpha=1$ and 1.25
and  the gas density between few times $10^{4-6}$ cm$^{-3}$, 
the magnetic field strength is constrained to values between $\sim0.5$ and 1 mG, respectively.

A  widely accepted alternative model
argues that the 6.4 keV  FeI K$\alpha$ line emission 
results  from  irradiation of molecular clouds near the Galactic center 
by a hypothetical transient source associated with Sgr~A*, which was active 
about 400 and again 100 years ago$^{21}$.  
The variability of 6.4 keV line emission on a short time scale provided a strong evidence 
in support of this hypothesis. However, the evidence for nonthermal emission from molecular clouds,  
as presented here,  also 
predicts time variability in the context of the cosmic ray picture$^{42}$. 
A supernova event  expanding inside the cloud 
could change not only the ionization characteristics of the molecular gas in G0.13--0.13 but also 
causes  the  observed 6.4 keV time variability. 
Thus, the time variability of K$\alpha$ line emission need not be clear support for the 
X-ray   irradiation  model.

\subsection{Cosmic Ray  Heating of Molecular Gas}
 
To estimate the temperature of molecular gas subject to high levels of cosmic-ray ionization, we extended our previous 
estimates of the cooling rate$^{42}$ 
 to higher densities, using the fitting formula for cooling by 
rotational transitions of CO, H$_2$, and H$_2$O provided by Neufeld \& Kaufman (1993) and Neufeld, Lepp \& Melnick (1995).  We 
also included an LVG calculation of the contribution by OI fine-structure lines, using the atomic data and collision rates in 
the Leiden Atomic and Molecular Database$^{49}$. 
  We adopted an abundance of CO relative to H$_2$ of 
$2.8\times10^{-4}$, twice the values adopted by Neufeld, Lepp \& Melnick (1995) to reflect the higher metallicity of the gas 
in the CMZ.  The abundance of OI and H$_2$O is temperature-dependent because of the conversion of OI and O$_2$ to H$_2$O by 
neutral-neutral reactions at temperatures above 200 K, so we assume  twice the chemical equilibrium abundances 
of OI and H$_2$O found by Neufeld, Lepp \& Melnick (1995): OI$/$H$_{2} = 2\times10^{-4} / (1+(T/220\,\mathrm{K})^{14})$, and 
H$_2$O$/$H$_2 = \mathrm{dex}( -6.222 + 2.831 / ( 1 + (245 \mathrm{K}/T)^{14} ) )$.  We adopted a cloud column of $2\times 
10^{23}$\,H$_2$\,cm$^{-2}$ and a line FWHM 20\,km\,s$^{-1}$.

The resulting total cooling rate per H$_2$ molecule is plotted in Figure 7b for representative H$_2$ densities of $10^3$, 
$10^4$ and $10^5$\,cm$^{-3}$. The cooling rates on the left hand axis are mapped to the right-hand axis showing the cosmic ray 
ionization rate that would supply heat at the same rate, assuming that each ionization is associated with the deposition of 
12.4\,eV of heat$^{42,52}$. 
The dip in the cooling rate at $T\approx250$\,K is due to the increasing  importance of OI and H$_2$O 
cooling at higher densities and the switch from OI to H$_2$O at this temperature.  
We conclude that an ionization rate of $10^{-14}$ to 10$^{-13}$\,s$^{-1}$ would yield gas 
temperatures in the range 50--200 K, the upper range consistent with the temperatures estimated using LVG shown in Figure 
5b,c.

%The contribution of protons can also be 
%significant for fixed flux measured from radio observations. 
%The contribution of 
%protons in the cosmic ray ionization rate for three different spectral index $\alpha=-0.5, -1$, and 
%$-0.25$. The bottom and top curves show the total cosmic ray electron density and magnetic field 
%strength, respectively, to yield a given synchrotron emissivity.  
%These plots  suggest that as we increase the fraction of protons to electrons 
%from 0 to 100, the cosmic ray ionization rate increases by a factor of few. 

\subsection{High Velocity Dispersion of Molecular Clouds}

High cosmic-ray fluxes in molecular clouds affect star formation by
heating the gas and increasing its ionization fraction. 
 Higher cloud
temperatures increase the Jeans mass, potentially changing the IMF,
while high ionization increases magnetic coupling to the cloud
material, reducing ambipolar diffusion and increasing the time
taken for gravitationally unstable cores to contract to the point that
they overwhelm their magnetic support. Another consequence of increased ionization
is the therefore reduced damping of
MHD waves, contributing to sustaining Alfvenic velocity
fields within the clouds, which may assist in explaining the observed
high velocity dispersion of molecular clouds in the nuclear disk (e.g.
 Oka et al. 1998; Martin et al. 2004).
In a weakly ionized medium, waves with
frequencies $\omega \sim kv_A$ below the collision frequency of neutral
particles with ions, $\nu_{ni} = n_i <\sigma v>$,  are damped
on a time scale $2 \nu_{ni} / \omega^2 $ $^{53,54}$
directly proportional to $n_i$ which is 
increased by a factor of $\sim100$ compared to 
the Galactic disk for $\zeta\sim10^{-13}$ s$^{-1}$. 
The
power input required to maintain wave motions on a given scale is
reduced by the same factor.

%We showed  in equation (3) of Yusef-Zadeh et al. (2012), 																																											
%\begin{equation}
%\zeta = 1.6\times10^{-13}\,  \left(\frac{\rm I_{\nu}\,   \nu^{(p-1)/2}}{(p-1)\,  L\,  \rm
%B^{(p+1)/2}}\right)
%    \ut s -1 \ut H -1
%\end{equation}

%I looked into protons.  At MeV energies they are 20 times more ionizing than electrons, and their loss time is correspondingly 
%shorter. I've attached a rough plot of the loss time for protons and electrons.

\subsection{Molecular Line Spectra toward G0.13--0.13}

%Given the high value of cosmic 
%ray ionization rate found throughout the Galactic center, the chemistry of the gas is considered to be driven by  
%cosmic rays. 

Figure 8a-d show line emission from 
16 molecular lines  toward  four  positions in G0.13--0.13 
corresponding to  A to D in
Table~1, respectively. The spectra of these 
pointed observations
show a rich chemistry similar to that seen throughout the 
CMZ.  We focus only on five molecular lines 
HCN (1--0), HCO$^+$(1--0),  HNC(1--0), N$_2$H$^+$(1--0)  and SiO (2--1), as 
Table~2  show the peak velocities and peak intensities in T$^*_A$ of all 
five spectral lines.   
Positions  A and D fall in the  E and NW boxes (see Fig. 5a). 
As a demonstration, 
the  spectrum of
these five molecular species 
toward  position A  is shown in Figure 9. 
The strong velocity component associated with G0.13-0.13 is centered around 60 \kms\,  (see Table 2). 
We also notice  an additional weak velocity component between -50 \kms  and  
0 \kms. 

% We assumed that the  
%component is not  associated with G0.13-0.13. 

Using RADEX\footnote{http://www.sron.rug.nl/~vdtak/radex/index.shtml} (van der Tak et al. 2007)$^{55}$, 
we derived the column density of each  molecular species   to match the 
observed intensity  
with the assumption that molecular gas temperature 
and  gas density  are  T$=200$ K and  10$^4$ cm$^{-3}$, respectively and 
the velocity dispersion is 30 \kms. 
Table 3 shows the log of
column density ratio of  HCN/HNC, HCO$^+$/N$_2$H$^+$ and SiO/N$_2$H$^+$ in columns 2-4 and 
the corresponding antenna temperature ratios 
in column 5-8 based on  deep Mopra observations at the 
 four positions A-D given column 1.  Calibration uncertainties of 
25\% have not been included in tables 2 and 3. 
We note that position C lies at the center of 
the  boot-shaped structure, cf. Fig.  2b,  where 
the 74 MHz emission peaks,   implying that  the cosmic ray ionization rate is higher than the other three positions. 

%peak intensity ratios of SiO(2-1) to HCN (1-0), HCO$^+$ (1-0), HNC (1-0) and N$_2$H$^+$ (1-0) 
%and found that the line ratios  are quite similar to each other for all four 
%positions except position C.  The line ratios for position C 
%are 0.12, 0.14, 0.28 and 0.27 and 
%are  lower than other three line ratios by a
%factor 1.5 to 2.  

%We considered the evolution of a diffuse cloud with initial density of 10$^6 \u cm -3$, visual 
%extinction of $\sim300$ mags, gas temperature 10 K, and standard cosmic ray ionization rate.  This cloud 
%is then chemically evolved with the assumption that cosmic ray ionization rate  $10^{-14}$  s$^{-1}$ 
%heats the  gas to a temperature 200 K and a dust temperature of 30 K. 
%The difference between the gas and dust temperature results 
%from the mechanism in which ices are evaporated from dust grains, i.e. the sublimation, by cosmic rays as 
%a function of time rather than by thermal heating.

\subsection{Chemical Modelling}

To examine the effect of high cosmic-ray ionization rates on chemistry,
we make use of a time-dependent gas-grain chemical model, UCL\_CHEM$^{31}$
 to qualitatively investigate the behavior of the observed species. The UCL\_CHEM is a gas-grain time-dependent model; for the 
purpose of this paper it is used as a two-stage model: {Phase I follows the free fall collapse of a diffuse (10$^2$ cm$^{-3}$) 
gas to a denser state (where the final density is a free parameter), while Phase II follows the chemistry as the gas and dust 
warm up} due to either the increase of temperature and/or an enhanced cosmic ray ionization rate (Phase II).  Apart from the 
very initial diffuse state (where the temperature is about 100K), during the collapse in {Phase I} the (coupled) gas and dust 
temperature remains constant at 10 K and atoms and molecules collide with, and freeze on to, grain surfaces. The advantage of 
this approach is that the ice composition is {\it\bf{not}} assumed but it is derived by a time-dependent computation of the 
chemical evolution of the gas--dust interaction process which, in turns, depends on the density of the gas. Hydrogenation 
occurs rapidly on these surfaces, so that, for example, some percentage of carbon atoms accreting will rapidly become frozen 
out methane, CH$_4$. The justification for a constant 10 K during most of Phase I is two-fold: first, in order for a 
diffuse gas to collapse under gravity and form a dense core, the temperature must remain low.  Second, it is believed that 
methanol must form on icy mantles (and this implies an efficient freeze out of CO which can only occur for dust temperatures 
less than 20K) since experiments show that it can not form in the gas phase$^{56}$ (e.g.,  Geppert et al. 2006). 
In Phase II {(which 
effectively represents the observed phase)} the gas and dust temperature are decoupled; the temperature of the gas is varied 
from 50 to 200 K, while the dust temperature was kept to 20-30K. We employ the reaction rate data from the UMIST astrochemical 
database, augmenting it with grain-surface (hydrogenation) reactions$^{31,57}$
In both Phases nonthermal desorption is also considered$^{58}$. 

% as in Roberts et al. (2007).

Figure 10 shows the abundance ratios of selected species as a function of time for
a subset of our model grid where the final density was varied from
10$^4$ cm$^{-3}$ to 10$^6$ cm$^{-3}$ and the cosmic ray ionization rate
from 10$^{-15}$ s$^{-1}$ to 10$^{-13}$ s$^{-1}$.  The gas temperature in
Phase II was assumed to be $\sim200$K.  Note,  however,  that while
molecular ratios are in principle powerful tools in constraining the
physical and chemical characteristics of a cloud, the assumption that
all the selected species arise from the same region may well be
incorrect. Nevertheless, we assume that the selected species
are co-spatial  and consider
a comparison between  the theoretical column density ratios with
those  derived from  our observations (see Table 3).
The  [HCN/HNC] (black) and [HCO$^+$/N$_2$H$^+$] (green)
values, as drawn on the top right panel in dashed lines, 
agree with models of low density ($10^4$ cm$^{-3}$) and high cosmic
ray ionization rate $\zeta=10^{-13}$ s$^{-1}$ H$^{-1}$.
We note an increase in the gas
density and or a decrease in the cosmic ray ionization rate
does not change the abundance ratio [HCN/HNC] but
increases the HCO$^+$/N$_2$H$^+$ (green) ratio to values well
above the observed range of 0.005 to 0.19 of Table 3.
The cosmic ray ionization rate,  inferred independently
from this model,  is remarkably
similar to that estimated from FeI K$\alpha$ line measurements (see Fig. 7a).
In addition, the gas density
predicted  from this  chemical model is consistent within the range of values inferred from LVG models, as shown in 
Figure 5b,c. 

%However, the column density estimated from  SiO 6-5 line inetnsity 
%toward position A  is 5$\times10^{15}$ cm$^{-2}$. Thus, 
%the columnn density ratio 
%[SiO(6--5)/N$_2$H$^+(1--0)]\sim0.8$ is closer to the value of $\sim1$
%predicted in Figure 11.  Future multi-transition observations should be able to 
%test if the high abundnace of SiO is  consistent with chemical modelling presented 
%here.  

It is known  that Galactic center  molecular clouds have 
high abundance of SiO$^{59,60,61}$. 
In one estimate, Minh et al. (1992) find
the SiO abundance relative to molecular hydrogen to be between  $\sim10^{-7}$ and  
$10^{-8}$. In other studies, 
Martin Pintado et al. (1997) also  find a high abundance of SiO
$\sim10^{-9}$  toward Galactic center molecular clouds. 
The observed [SiO/N$_2$H$^+$] abundance ratio ranges between 0.27--0.44 and can not be matched by any
model at late enough times that chemical equilibrium has been reached. The models predict even higher 
fractional abundance of [SiO/N$_2$H$^+$]. 
In our models,  we used an initial atomic abundance for Si depleted by a factor of 100 with respect to solar, 
as this represents the lower limit of 
measured initial abundances in a large sample of molecular clouds$^{62}$.  
However, all initial abudances$^{63}$ 
are then scaled to match a metallicity twice the solar value. 
Note that the abundance of
SiO derived by our models is
consistent with the fact that
with high cosmic ray ionization rates the icy mantles are released and therefore Si
is released in the gas phase. It is therefore puzzling to observe such a
low fractional abundance in a gas where we expect cosmic ray ionization rates to be enhanced.
There are three possibilities that could account for the low fractional abundance of
[SiO/N$_2$H$^+$] compared to that
predicted by chemical model.  
One is that the medium is quite turbulent and hence chemical
equilibrium is never reached and the gas is constantly being recycled; then the best
solution must be found at earlier times ($< 10^4$ years). In this scenario  all ratios can
be qualitatively matched for the range of densities 10$^{4-5}$ cm$^{-3}$ and the whole range
of cosmic ray ionization rates investigated and no further constraints can be given by the
models. The other is that the initial atomic abundance of Si is low. Although we already
employ, in  models shown in Figure 10, an abundance of the initial
Si 1.64$\times10^{-7}$  that is up
to a factor of 100 depleted with respect to solar$^{64}$. 
It   could be that in these environments even more than a factor of 100  is depleted.
Lastly, it is possible that our derivation of the column
density from SiO (2-1) intensity is not accurate, e.g. optically thick. 
Thus, the intensity of high rotational
transitions are needed to determine accurately the SiO column density.
Additional detailed work is needed to sort out the discrepancy  between the observed 
and modeled  [SiO/N$_2$H$^+$] abundance ratio.  What is clear  from our chemical modeling, as in shocks and 
and PDRs, a high  SiO can also be generated by cosmic rays.

\subsection{Summary}

We showed that the molecular emission from a Galactic center  cloud 
G0.13-0.13,  a representattive molecular cloud in the CMZ, 
 coincides with 74 MHz and FeI K$\alpha$ line emission at 6.4 keV. 
This three-way correlation provides 
compelling evidence  that 
relativistic electrons are physically associated with 
the G0.13--0.13 molecular cloud. 
The   high cosmic ray electron ionization rate 
$\sim10^{-13}$  s$^{-1}$ H$^{-1}$  
is  responsible for the 
FeI K$\alpha$ line emission,  heating the gas clouds to high temperatures 
and maintaining  the velocity dispersion of clouds by increasing the wave 
damping time scale.   
One of the important conclusions of this study is
that LVG modeling 
of  multi-transition SiO observations  gave 
molecular gas densities $\sim 10^{4-5}$ cm$^{-3}$ and temperature 
$\sim100-1000$K. 
The variation of the spectral index of nonthermal emission from 
G0.13-0.13  was  shown to be unusually  large $\Delta\alpha\sim 1.5$. 
We suggested   that the expansion of  the G0.13--0.13 molecular cloud 
into   nonthermal vertical filaments 
accelerates  particles along the filaments$^{25,48}$,  
thus producing 
a young population of electrons with an  unusually flat  energy spectrum. 

Evidence for FeI K$\alpha$ line and  74 MHz synchrotron emission 
from  warm  molecular gas 
supports the view that  the Galactic center is a cosmic ray dominated region. We  explored the 
chemistry in a region subject to high levels of cosmic ray ionization  
and  showed that enhanced  SiO emission, 
widespread throughout Galactic center molecular clouds, can be produced in the context of cosmic ray 
interaction with molecular gas. 
The model to explain enhanced SiO, as well as NH$_3$ and  CH$_3$OH, emission  by cosmic ray driven chemistry 
is an alternative to the shock-driven  chemistry throughout Galactic center molecular clouds. 

%, two orders of 
%magnitude more than observed values.  We suggested  that  the initial abundance of Si in grains 
%must be low in order to be consistent with measured column density ratio SiO/N$_2$H$^+$.  
%The  cosmic ray driven chemistry is consistent with observations that there is  

\acknowledgements 
This research is supported in part by grants from the 
$Fermi$ Guest Investigator Program as well as the grant AST-0807400 from 
the NSF the National Science Foundation. The Caltech Submillimeter 
Observatory is operated by the California Institute of Technology under 
cooperative agreement with the National Science Foundation 
(AST-0838261). Any opinions, findings, and conclusions or 
recommendations expressed in this material are those of the author(s) 
and do not necessarily reflect the views of the National Science 
Foundation.

\newcommand\refitem{\bibitem[]{}}

\begin{figure}
\centering
\includegraphics[scale=0.6,angle=0]{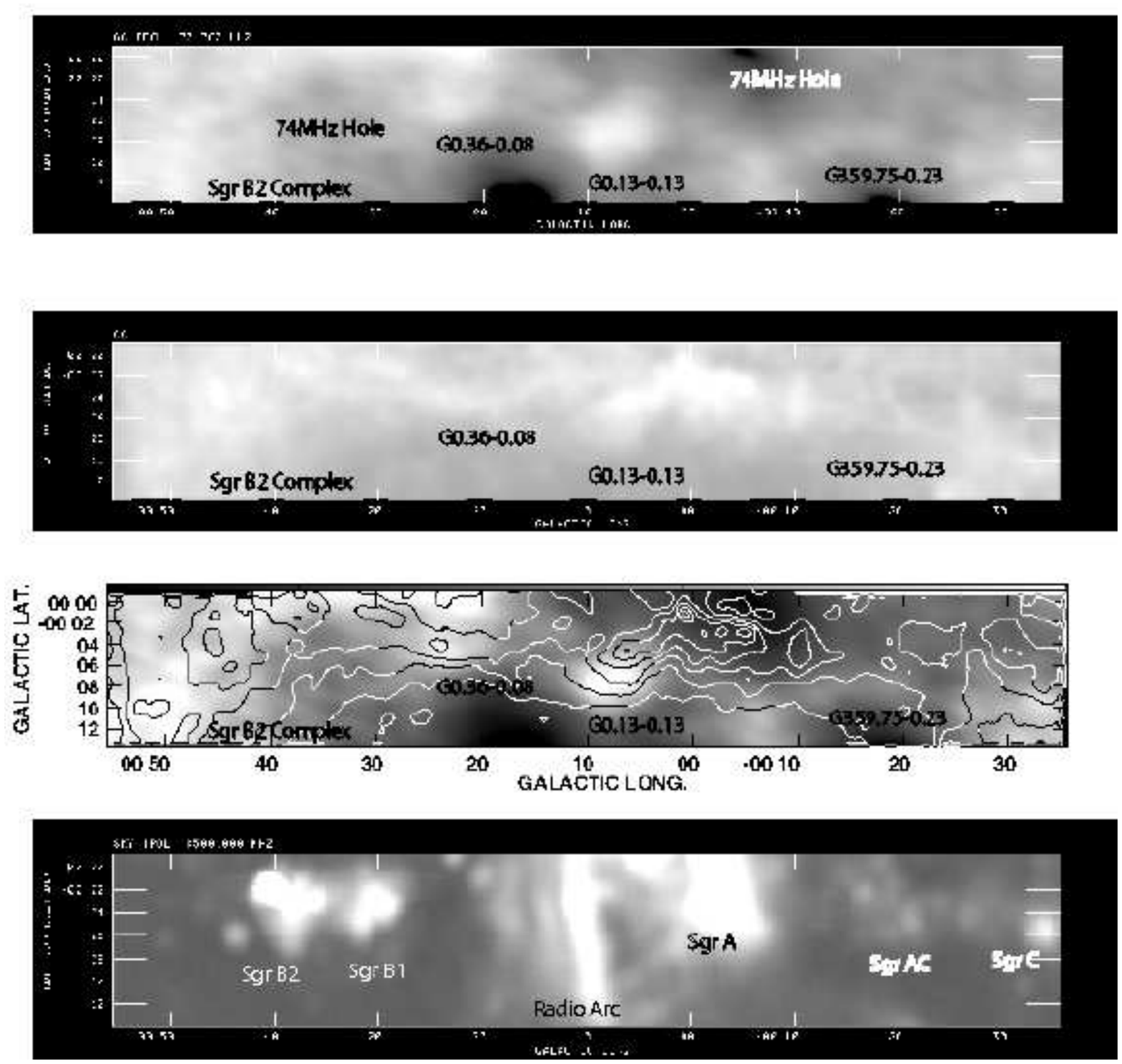}
\caption{
%\label{fig:20cm}
{\it (Top to Bottom)}
{\it (a)} 
A grayscale 74 MHz distribution  of the inner 
1.5$^{\circ}\times12'$ (Galactic coordinates $l\times b$)
convolved to  a resolution of $2'\times2'$ based on VLA observations. 
The 74 MHz flux range is between -1 and 4 Jy.  
{\it (b)} 
Similar to 
{\it (a)} 
except the distribution of CO (3-2) integrated over 
$\pm$200 \kms\, at  $34''$  resolution$^{48}$ with flux range 
100  and 2500 \kms\, K. 
{\it (c)} 
Contours of the CO (3-2) line emission shown in (b)
are superimposed on the 74 MHz image shown in {\it (a)}.
{\it (d)} 
Similar to {\it (a)} except observed with the GBT at 8.5 GHz with a resolution 
of 88$''$ $^{34}$. The flux range is between -5e-3 and 0.4 Jy. 
}
\end{figure}

\begin{figure}      
\centering           
\includegraphics[scale=0.3,angle=0]{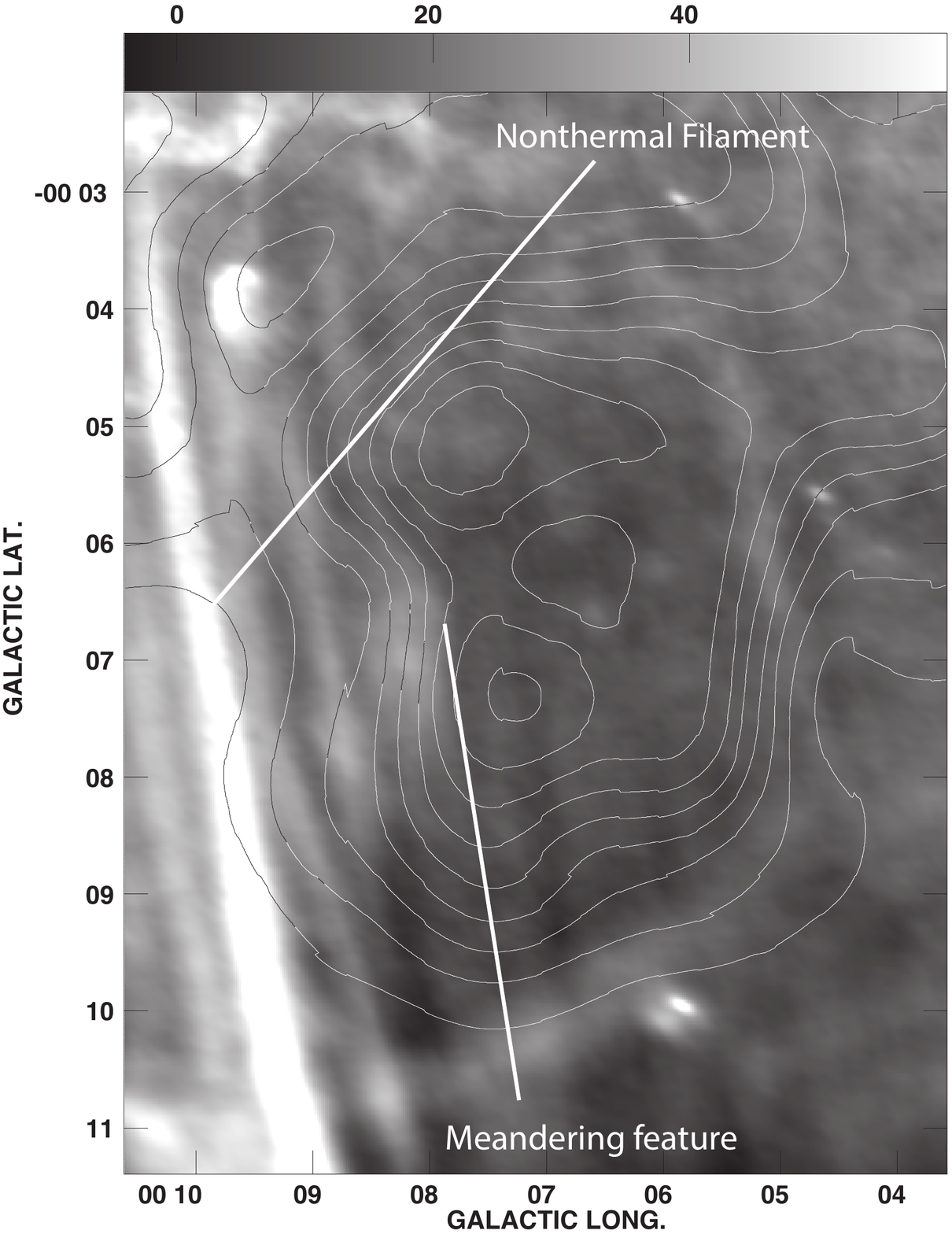}
\includegraphics[scale=0.3,angle=0]{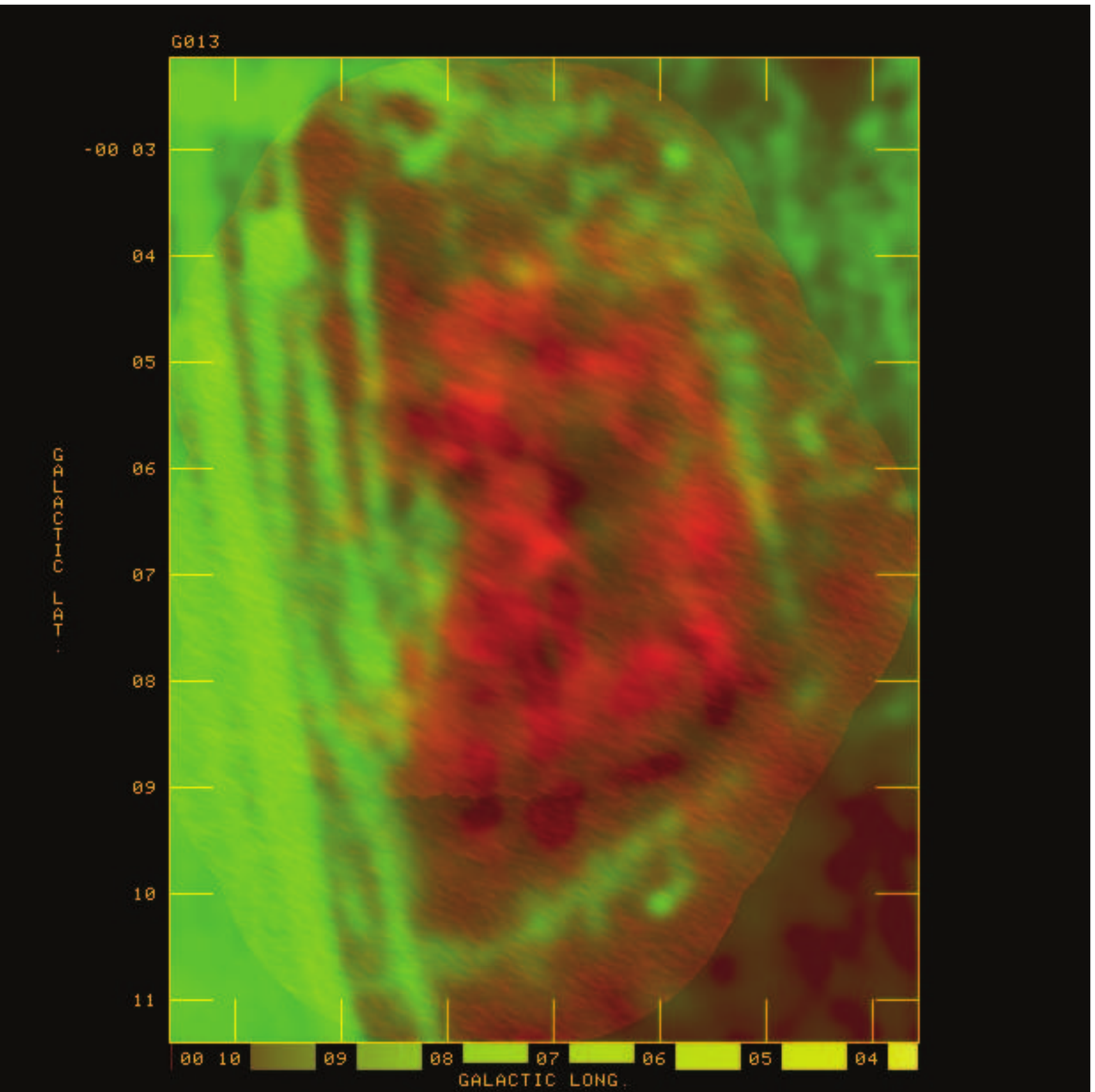}
\caption{
\label{fig:20cm}
{\it (a - Left)}
Contours of  CS (1-0) line emission integrated  between 0 and 50 \kms\,  
from the G0.13--0.13 molecular cloud with a
resolution of 45$''^{25}$  are superimposed on 
a 5 GHz  continuum image with a spatial resolution of 10.8$''\times5.5''$ (PA=-2.6$^{\circ}$). 
Contour levels are set at 2 to 10 \kms\,  K (T$^*_A$)
with 1 \kms\, K interval.  
{\it (b - Right)}
The distribution of CS (2-1) line emission (red) from G0.13--0.13 integrated over velocities
between --12.1 $< v_{LSR}<76.4$ \kms\, with a resolution of 
of  $3.1''\times2.6''$ (PA=20.5$^{\circ}$) is 
based on BIMA  observations. The 
extended  nonthermal filaments of the Arc at  1.4 GHz
with a resolution of  $10.7''\times10.1''$ (PA=28$^{\circ}$) are shown in green.
}\end{figure}

\begin{figure}
\center
\includegraphics[scale=0.3,angle=0]{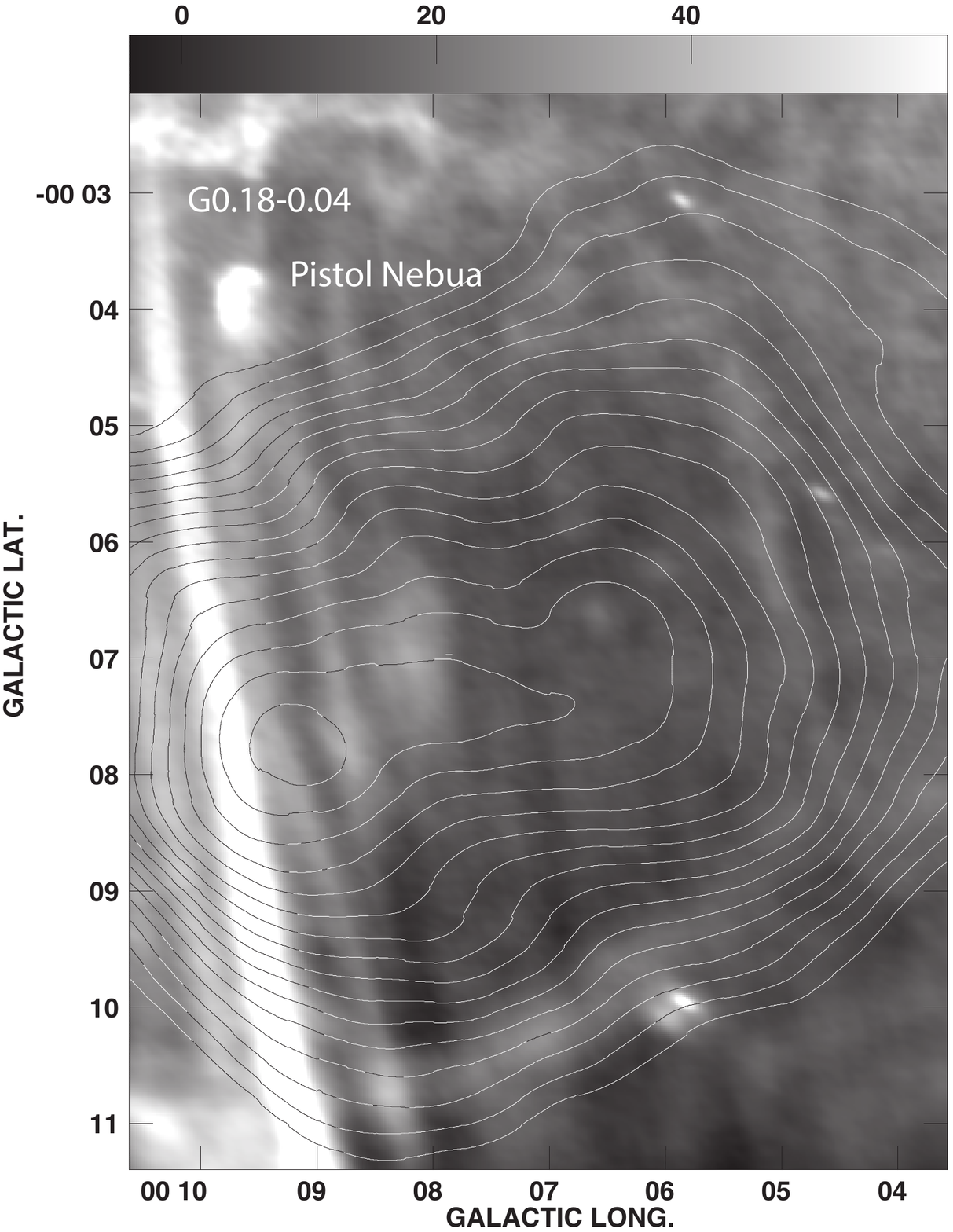}
\includegraphics[scale=0.3,angle=0]{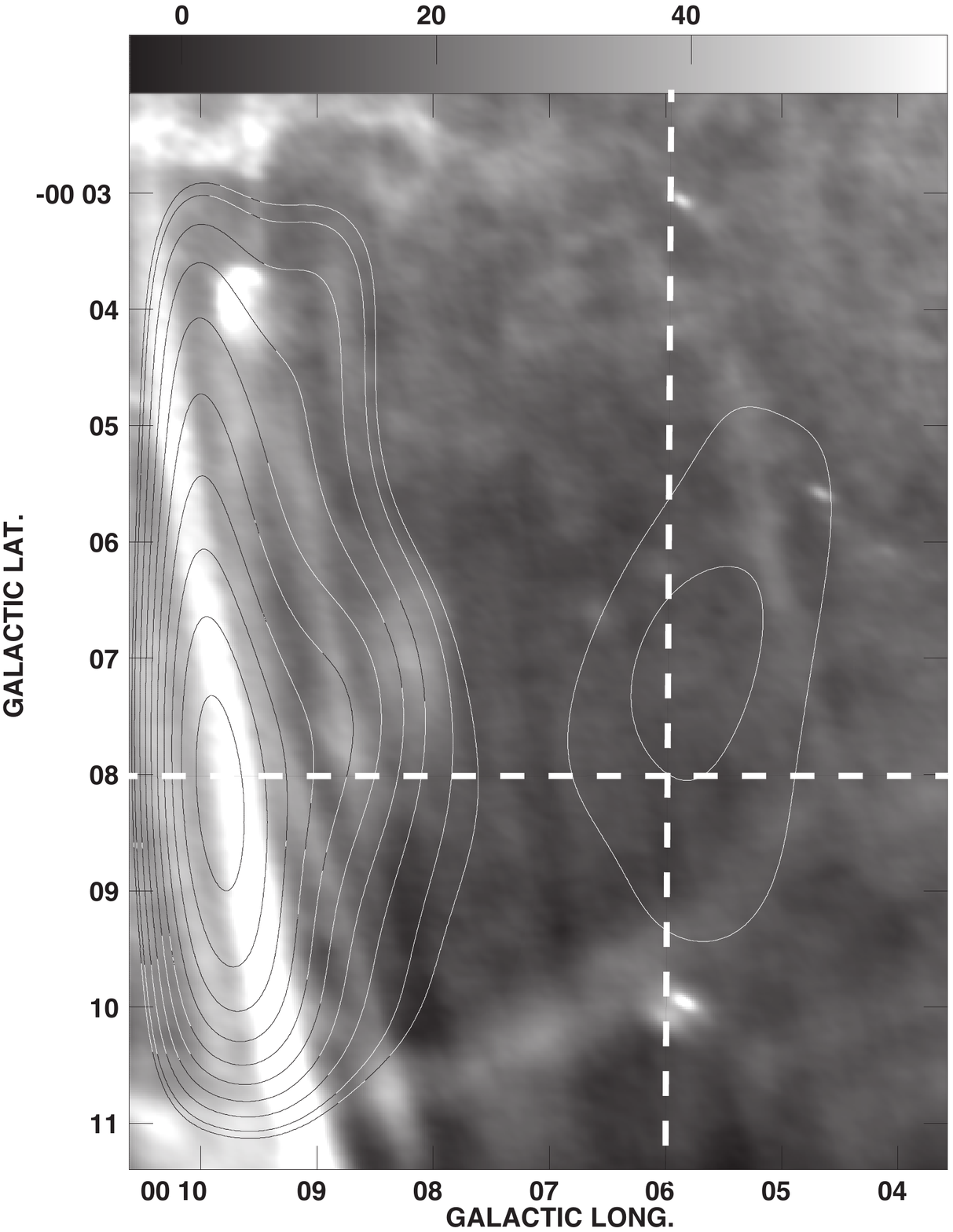}\\
\caption{
{\it (a - Top Left)}
Contours of 74 MHz emission  with a resolution 
122$''\times64''$ (PA=-$5^{\circ}$) with levels set at 0.8, 0.9,...2.3 Jy beam$^{-1}$ 
are superimposed on a grayscale image at 5 GHz.  
{\it (b - Top Right)} 
Contours of 327 MHz emission with levels 
(0.8, 0.9, ...2.3)$\times2$  are shown with the same resolution as in (a) and 
are superimposed on the 5 GHz image.
The white dashed lines show the location of cross cuts (see (d) below). 
{\it (c - Left)}
The spectral index distribution $\alpha$  between 74 and 327 MHz emission. The 74 and 327 MHz are
convolved to the same resolution and 
are background subtracted by 2 and 0.5 Jy, respectively. 
The bar at the bottom of the figure shows the color scale version of 
spectral index values ranging between -0.5 and 2 (see text). 
{\it (d-e - Right)}
Top two panels show background subtracted intensity profiles at 327 and 74 MHz (top) and the corresponding 
spectral index distribution between these two frequencies (below)  made from vertical 
latitude cross cut  at constant longitude l=$-6'$, as drawn on (b) with a vertical dashed line.   
Bottom two panels are the same as the top two panels except  that the cross cuts made 
horizontally at constant   b=$-8'$.  Cross cuts  are made on the 74 and 327 MHz images.
}
\end{figure}

% ******************************************
\setcounter{figure}{2} % reset figure counter to 1 so next figure is Figure 2 again.
% ******************************************
\begin{figure}[p] %  figure placement: here, top, bottom, or page
   \centering
\includegraphics[scale=0.3,angle=0]{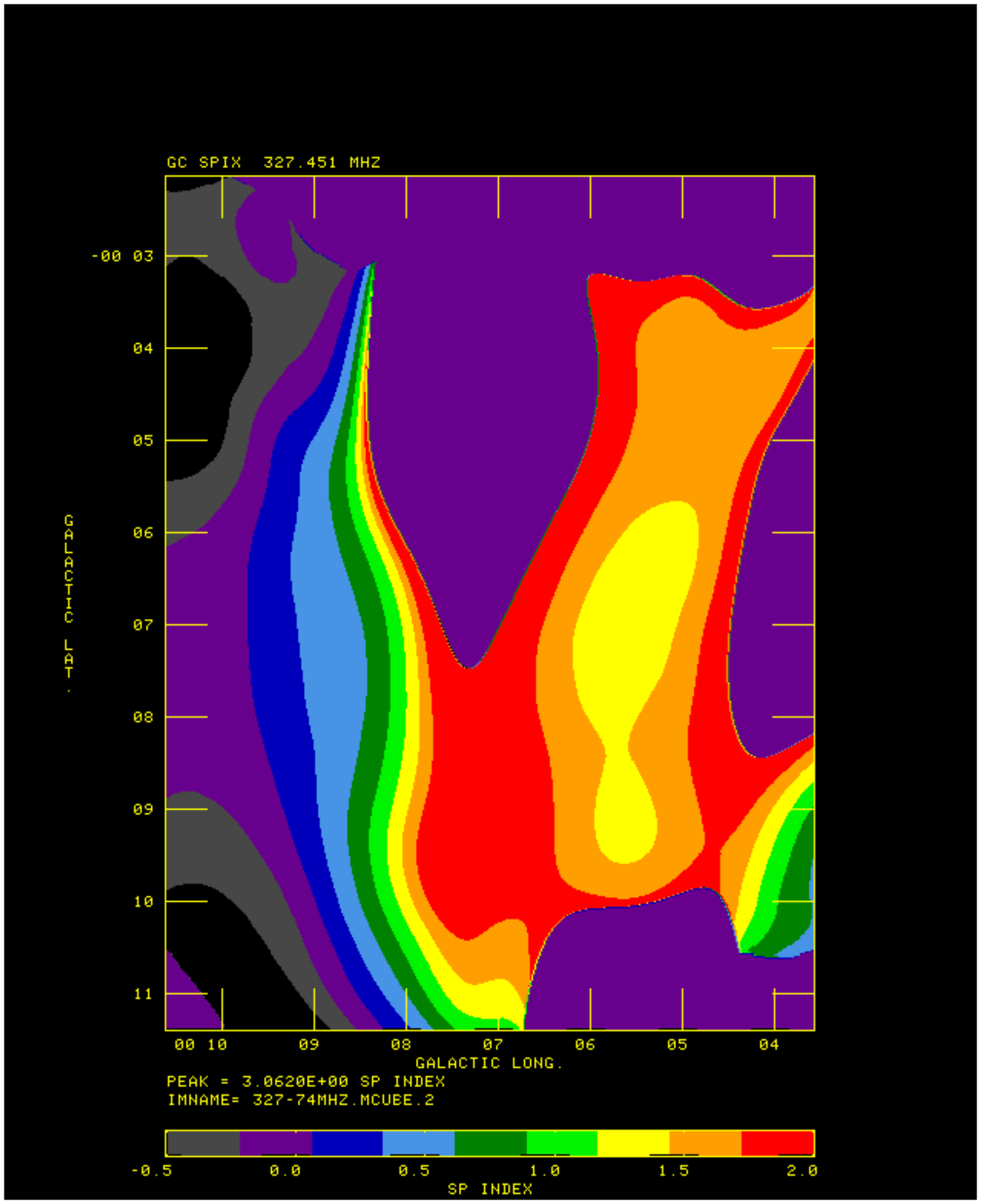}
\includegraphics[scale=0.4,angle=0]{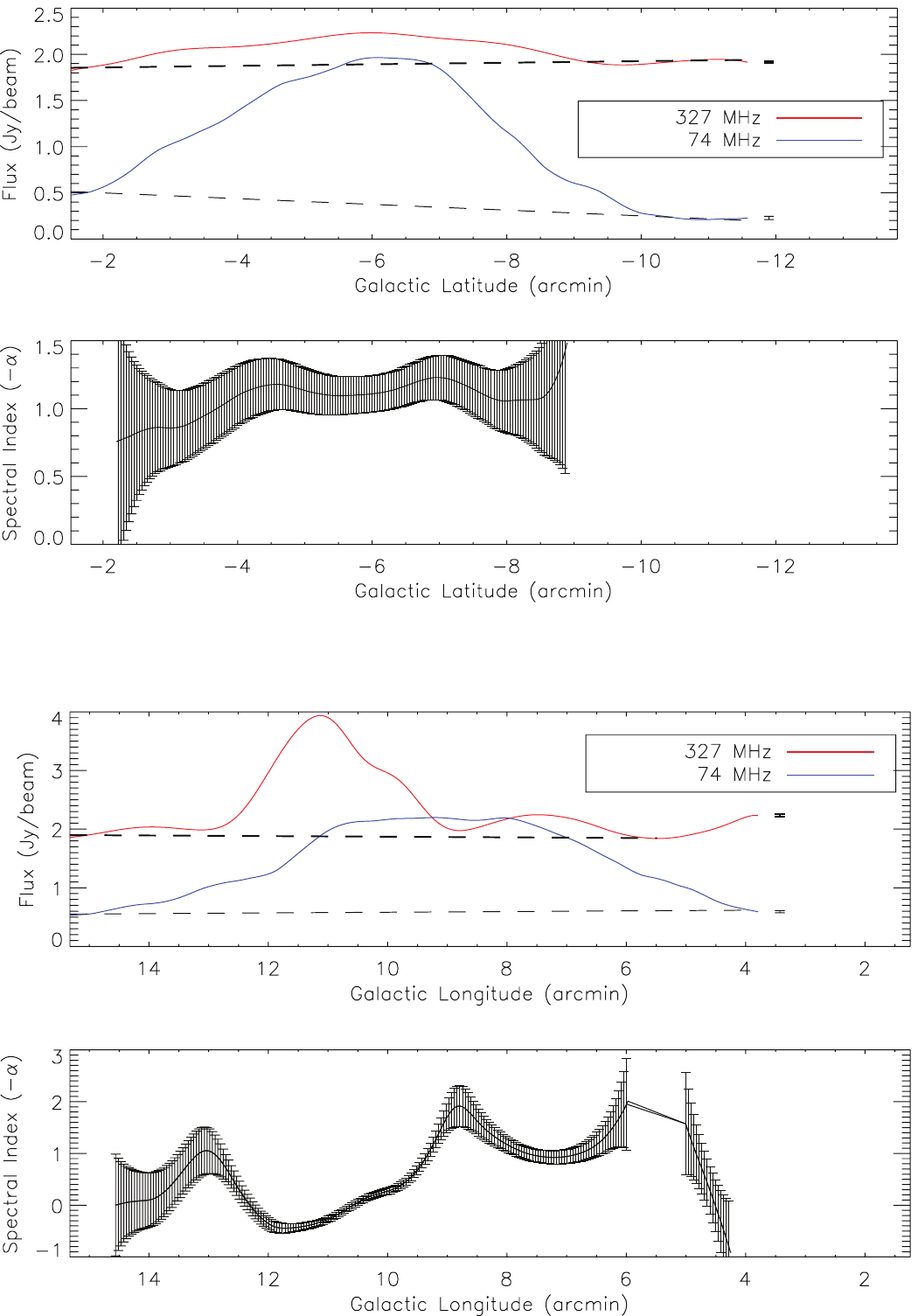}
   \caption{continued}
   \label{fig:example}
\end{figure}

\begin{figure}
\center
\includegraphics[scale=0.5,angle=0]{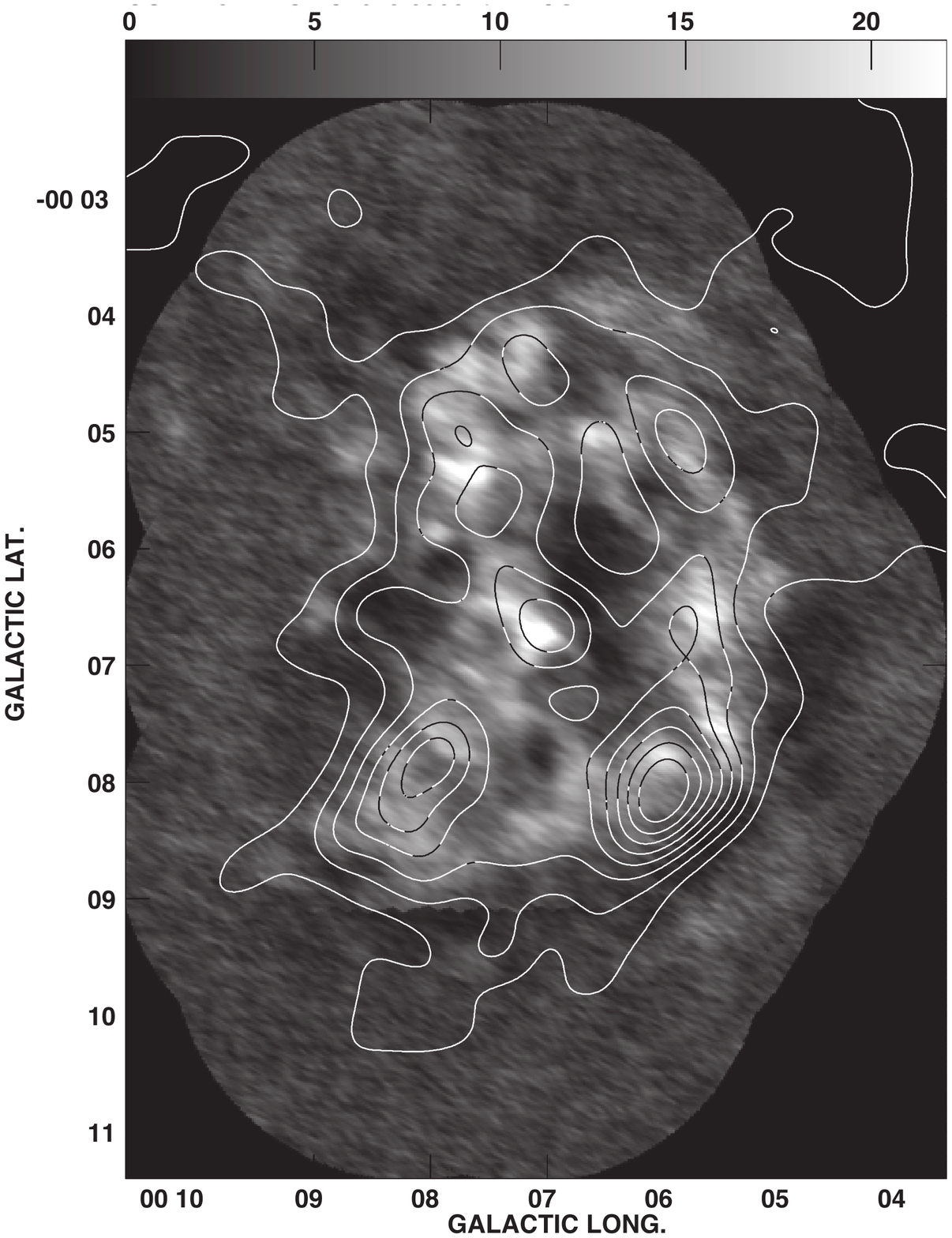}
\caption{
%\label{fig:20cm}
%{\it (a - Left)}
Contours of K$\alpha$ line emission at 6.4 keV line emission 
at (2, 3, ..., 9)$\times10^{-6}$ ph s$^{-1}$ cm$^{-2}$ beam$^{-1}$
are superimposed on a CS (1-0) line emission, as shown 
in Figure 2b.  CARMA image  shows emission between 6.25 and 6.5 keV 
which is convolved to a Gaussian beam of
30$''$. 
%{\it (b - Right)} 
%The same as (a) except that 
%contours of  EWs  of K$\alpha$ line emission with levels at 
%(1, 1.25, ...3, 3.5, 4, 5,...10)$\times100$ eV.
}\end{figure}

\begin{figure}
\center
\includegraphics[scale=0.35,angle=0]{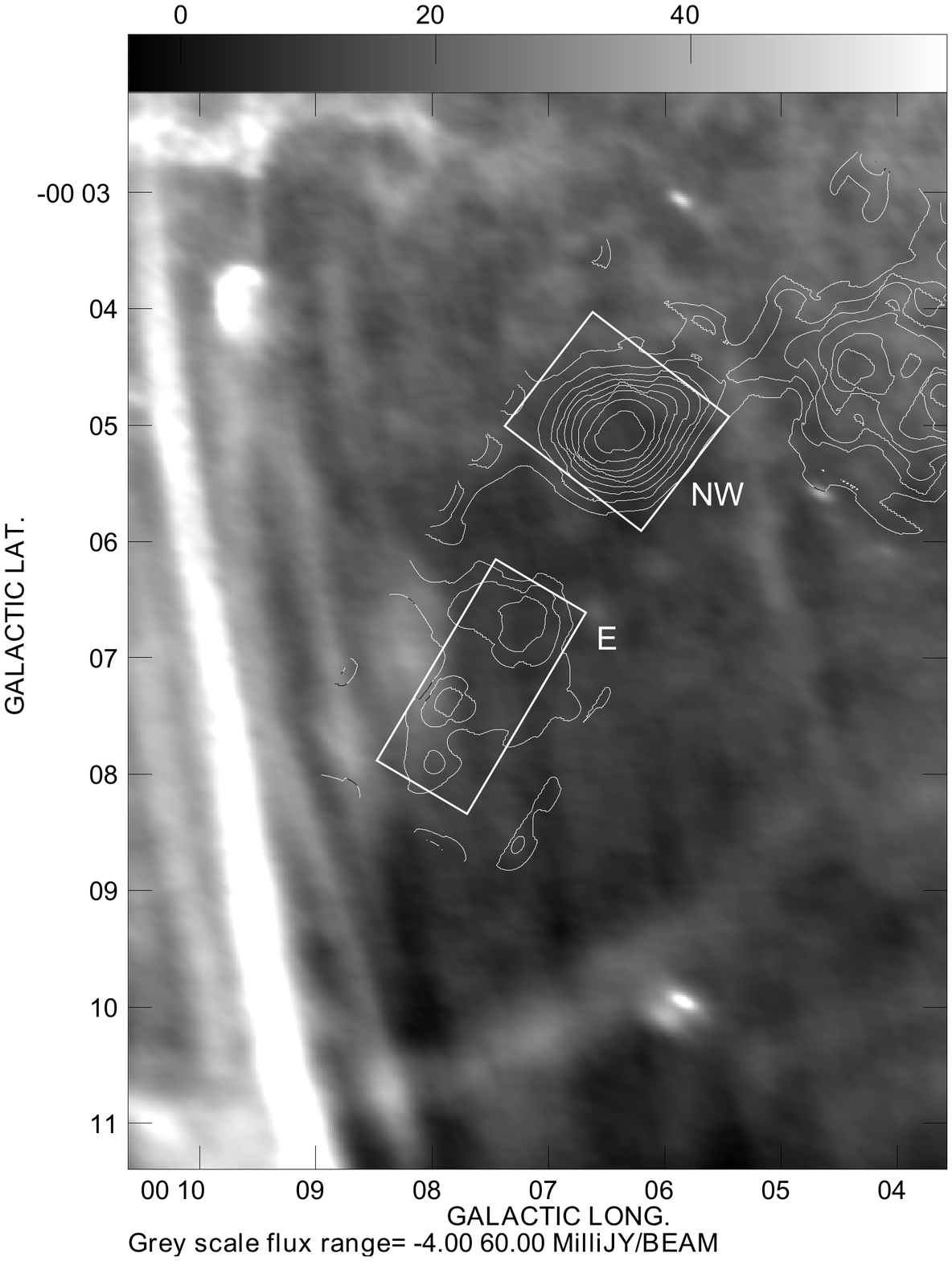}\\
%\vspace {10pt}
%\center 
%\includegraphics[scale=0.2,angle=0]{fig5b.pdf}
%\includegraphics[scale=0.2,angle=0]{fig5c.pdf}
\caption{
{\it (a)}
Contours of  SiO (5-4) line emission
integrated between -20 and 100 \kms
with levels at 8, 10, 12, 14, 16, 18, 20, 23 K \kms\, 
(T$_{\rm mb}$\,  velocity) are superimposed on a  5 GHz grayscale image. 
Two  rectangular boxes NW and E are drawn for LVG analysis.   
{\it (b)} 
The parameters of density n$_{\rm H}$ cm$^{-3}$  and  temperature T (K) using LVG model 
are presented.   
SiO (5-4), (6-5), (2-1)  line ratios are displayed in the bottom left corner  in red, green and black, respectively. 
The dashed lines show the corresponding calibration uncertainties 
of  the derived parameters at a level of 25\%.  
The SiO line intensity ratios  are extracted from  the  region presented 
by the E rectangular box, as drawn schematically on (a). 
{\it (c)} 
Similar to {\it (b)}  except that data  are 
from the NW box shown  in (a). 
{\it (d)} 
Top three  panels show the plots of 
 line ratios as function of logarithm of pressure (nT) based on a grid of models.  
The observed line ratio values with their corresponding errors are drawn horizontally (green) in 
solid and dashed lines, respectively. 
The bottom  panel shows  the best  $\chi^2$ 
fit to the grid of models.  
The vertical lines (green) are centered near minimum  $\chi^2$ 
giving log(nT) values  in the range of 6.4--6.9. 
}
\end{figure}

% ******************************************
\setcounter{figure}{4} % reset figure counter to 1 so next figure is Figure 2 again.
% ******************************************
\begin{figure}[p] %  figure placement: here, top, bottom, or page
   \centering
\includegraphics[scale=0.2,angle=0]{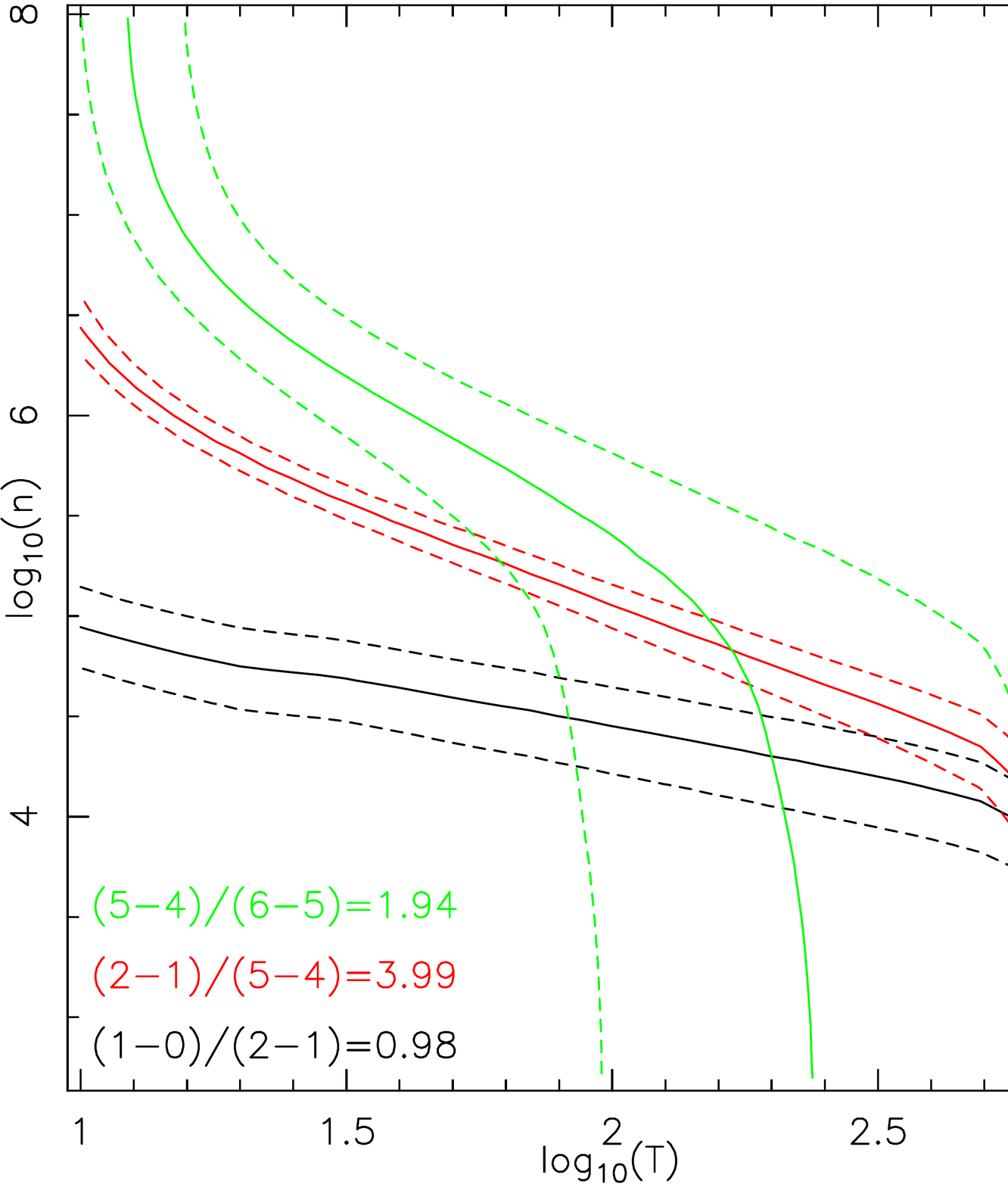}
\includegraphics[scale=0.2,angle=0]{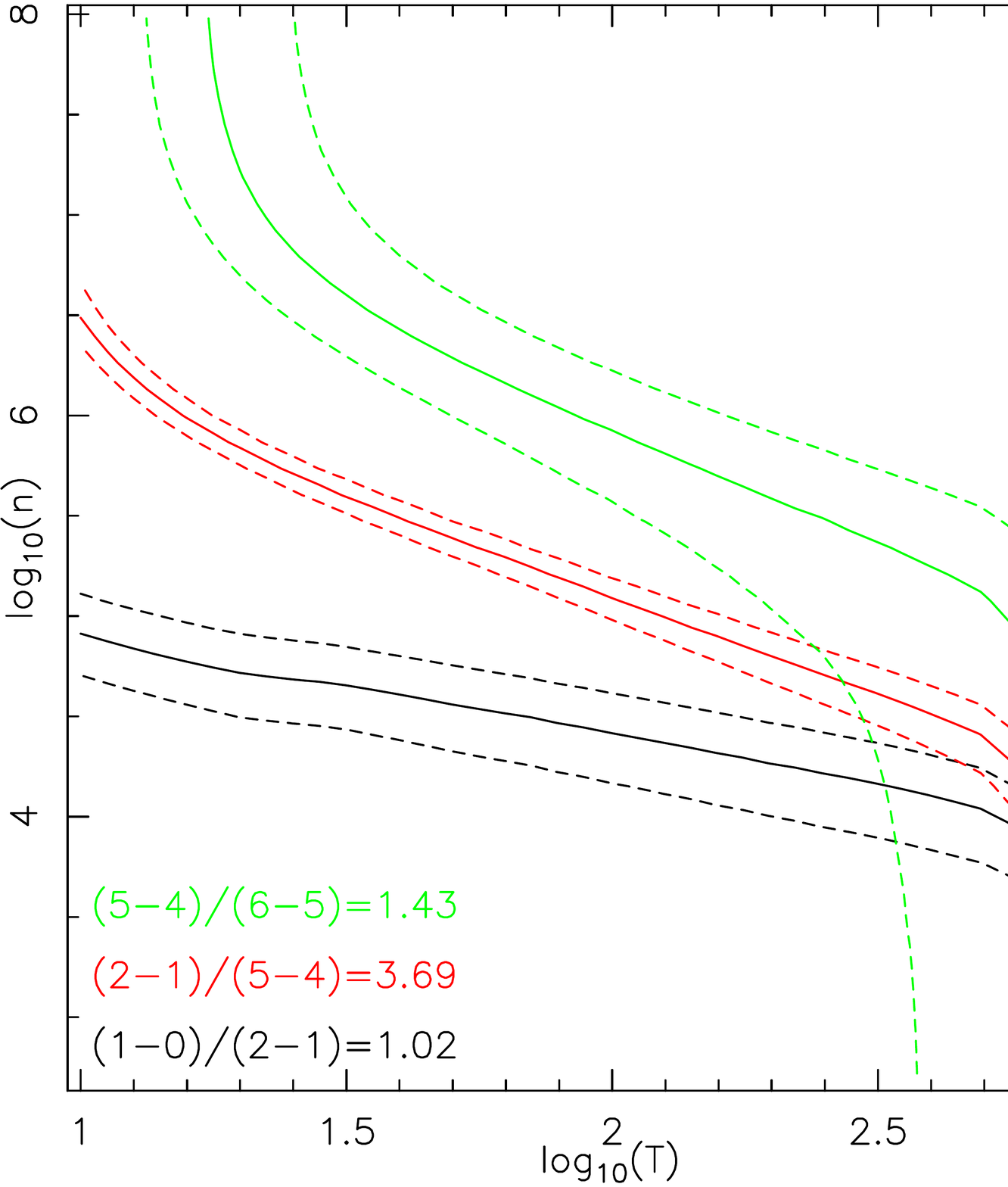}\\
\vspace {30pt}
\centering
\includegraphics[scale=0.4,angle=0]{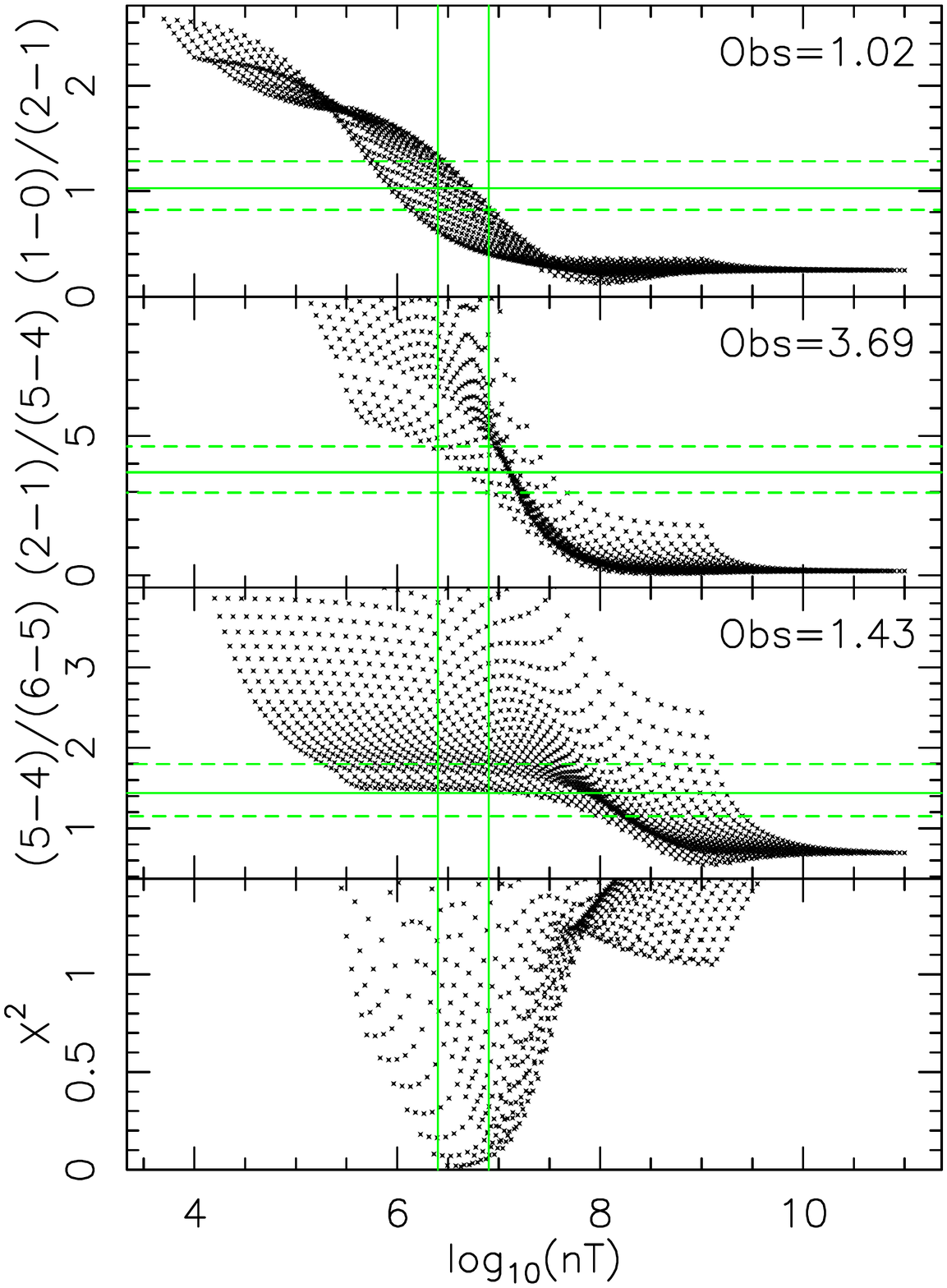}
\vspace {10pt}
\center 
   \caption{continued}
   \label{fig:example}
\end{figure}

\begin{figure}
\center
\includegraphics[scale=0.5,angle=0]{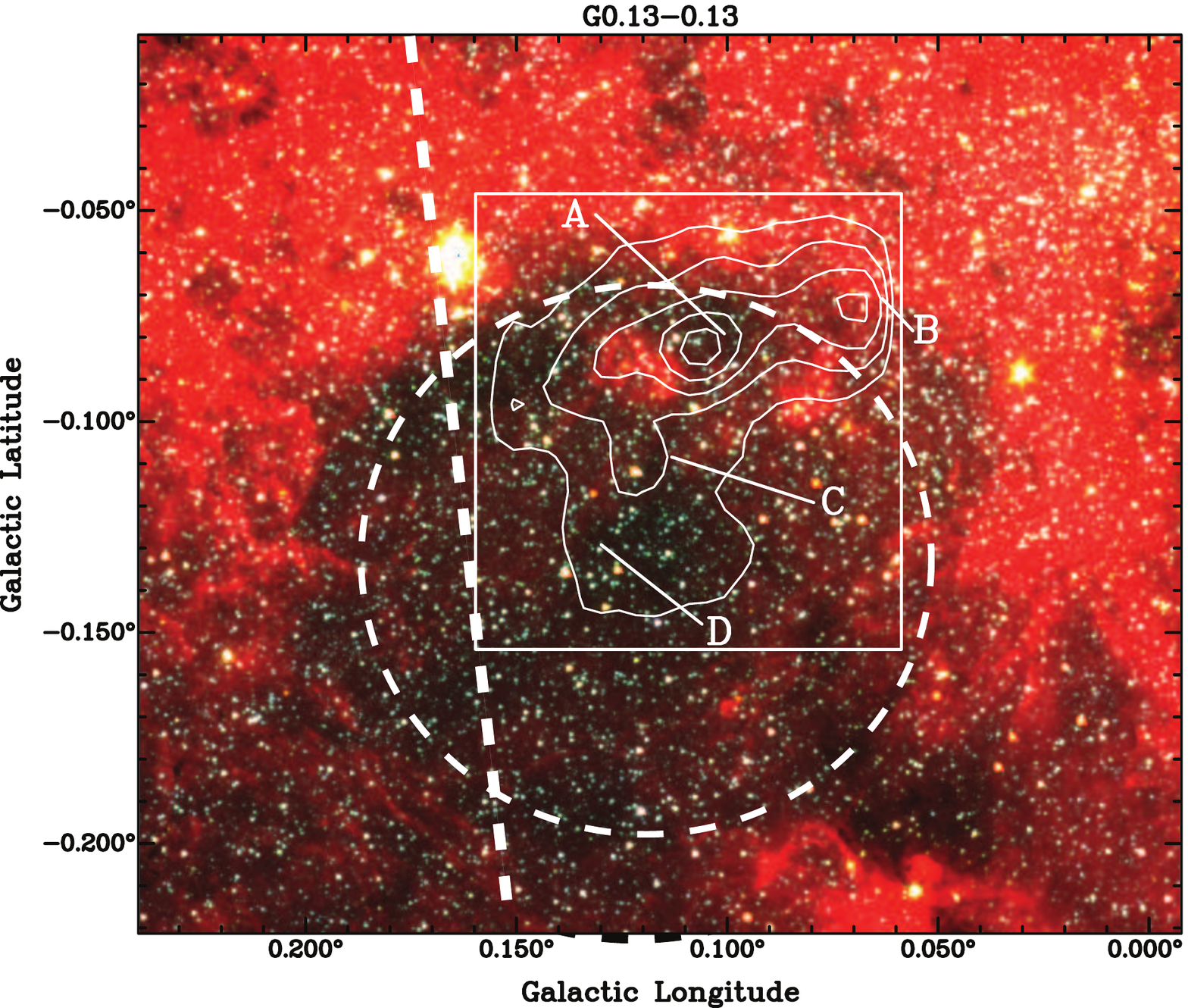}
\caption{Contours of SiO (2-1) line emission  based on Mopra observations 
with levels 2,  3.5,  5,  6.5,  8 K \kms\, are  superimposed on a three color Spitzer/IRAC 
image combining   Channels 4 (8$\mu$m) is  red), 2 (4.5$\mu$m) is green  and 1 (3.6$\mu$m) is blue. 
The temperature scale is in T$^*_A$.    The straight line in white shows the orientation of the 
nonthermal filament of the radio arc whereas the dashed circle marks roughly the size of the 
arc bubble. The positions  where molecular abundances are measured are drawn as A, B, C and D. 
The box indicated the region  mapped by Mopra.  
\label{fig:20cm}
}\end{figure}

\begin{figure} 
\center 
\includegraphics[scale=0.5,angle=0]{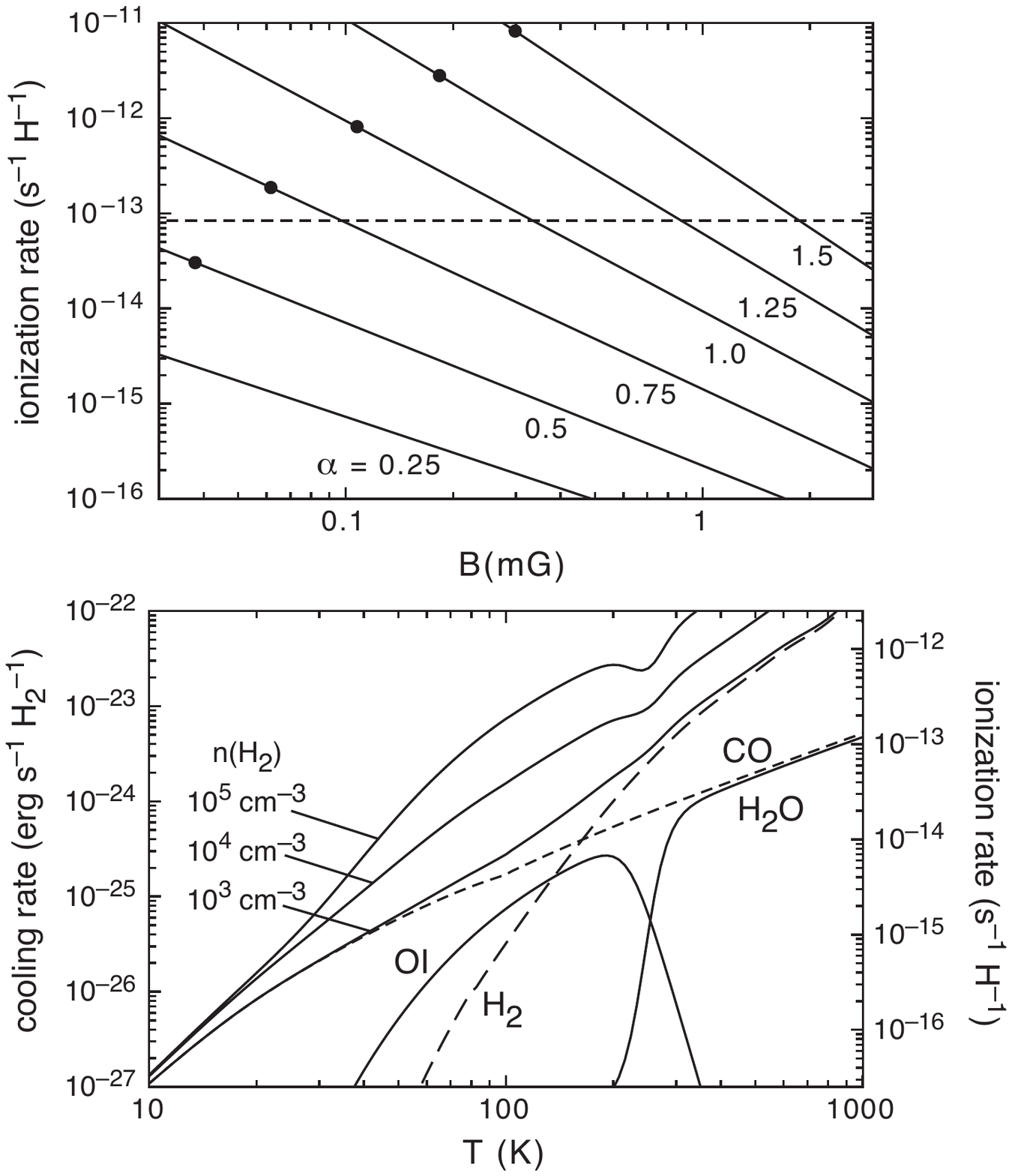}
\caption{ 
\label{fig:20cm} 
{\it (a - Top)} 
The variation of cosmic ray ionization rate as a function of magnetic field strength for different 
values of the spectral index of the radiation $\alpha$. 
The black dot on each curve gives the value at which the magnetic field and 
particle energies are in equipartition. 
{\it (b - Bottom)} 
Solid curves show the total cooling rate for diffuse molecular gas for H$_2$ densities of $10^3$, $10^4$ and 
$10^5$\,cm$^{-3}$. Rotational transitions of H$_2$, CO and H$_2$O, and by fine-structure transitions of OI  have been 
included; their individual contributions for $10^3$\,cm$^{-3}$ are indicated by the dashed curves. The right hand axis shows 
the ionization rate by cosmic-ray electrons needed to supply the corresponding heating rate (see text). The vertical axis
in both figures are  logarithmic.  
}
\end{figure}

%Solid curves show the total cooling rate for diffuse molecular gas for H$_2$ densities of 10$^3, 10^4$ and 10$^5$ cm$^{-3}$. 
%Only the dominant cooling, by rotational 
%transitions of H$_2$ and CO, has been included; these contributions for n(H$_2$) = 
%10$^3$ cm$^{-3}$ are indicated by the short-dashed and long-dashed curves, respectively. 
%The right hand axis shows the 
%ionization rate by cosmic-ray electrons needed to supply the corresponding heating rate, assuming that each ionization is 
%associated with the deposition of 12 eV of heat (see text).

\begin{figure}
\center
\includegraphics[scale=0.3,angle=0]{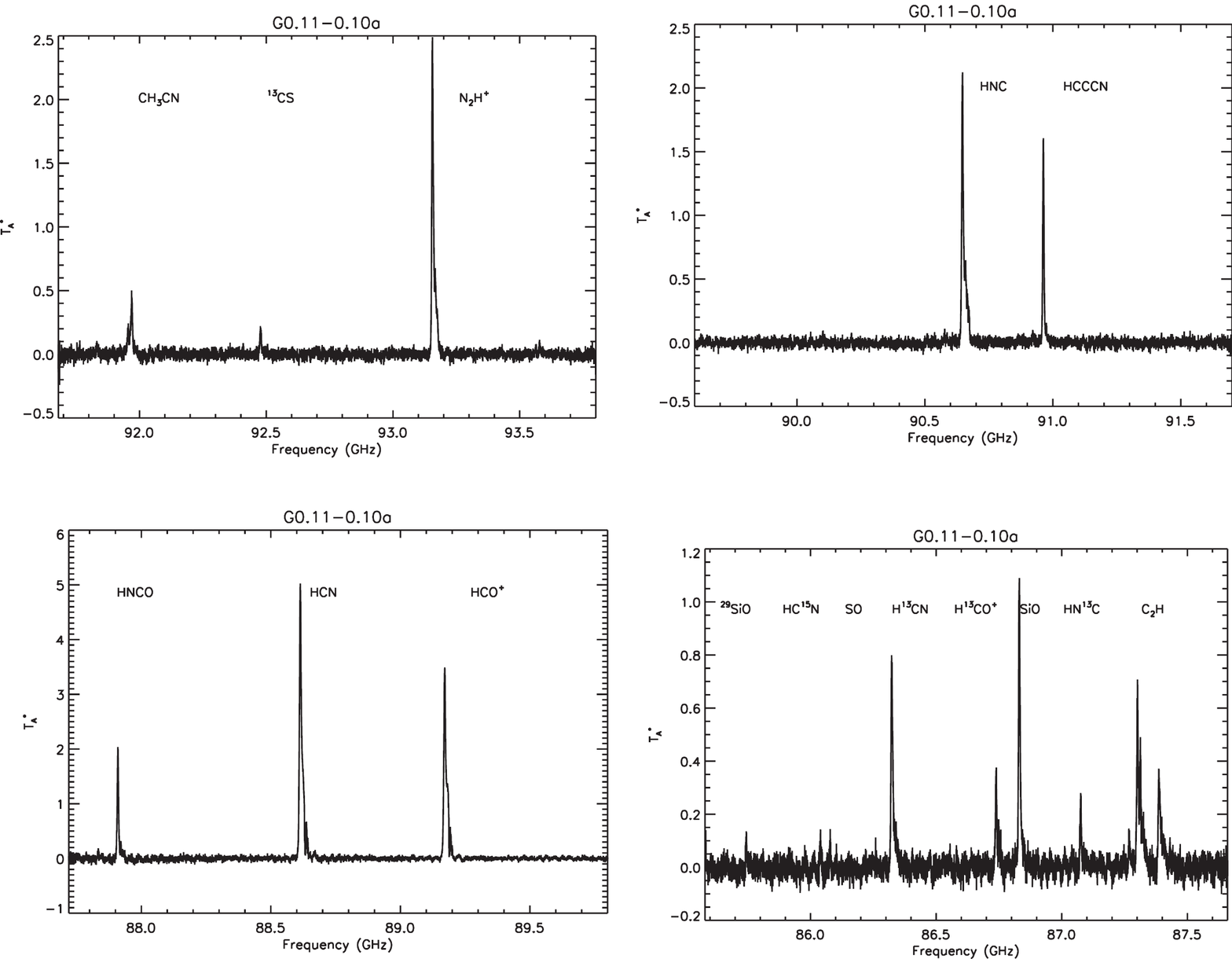}
\includegraphics[scale=0.3,angle=0]{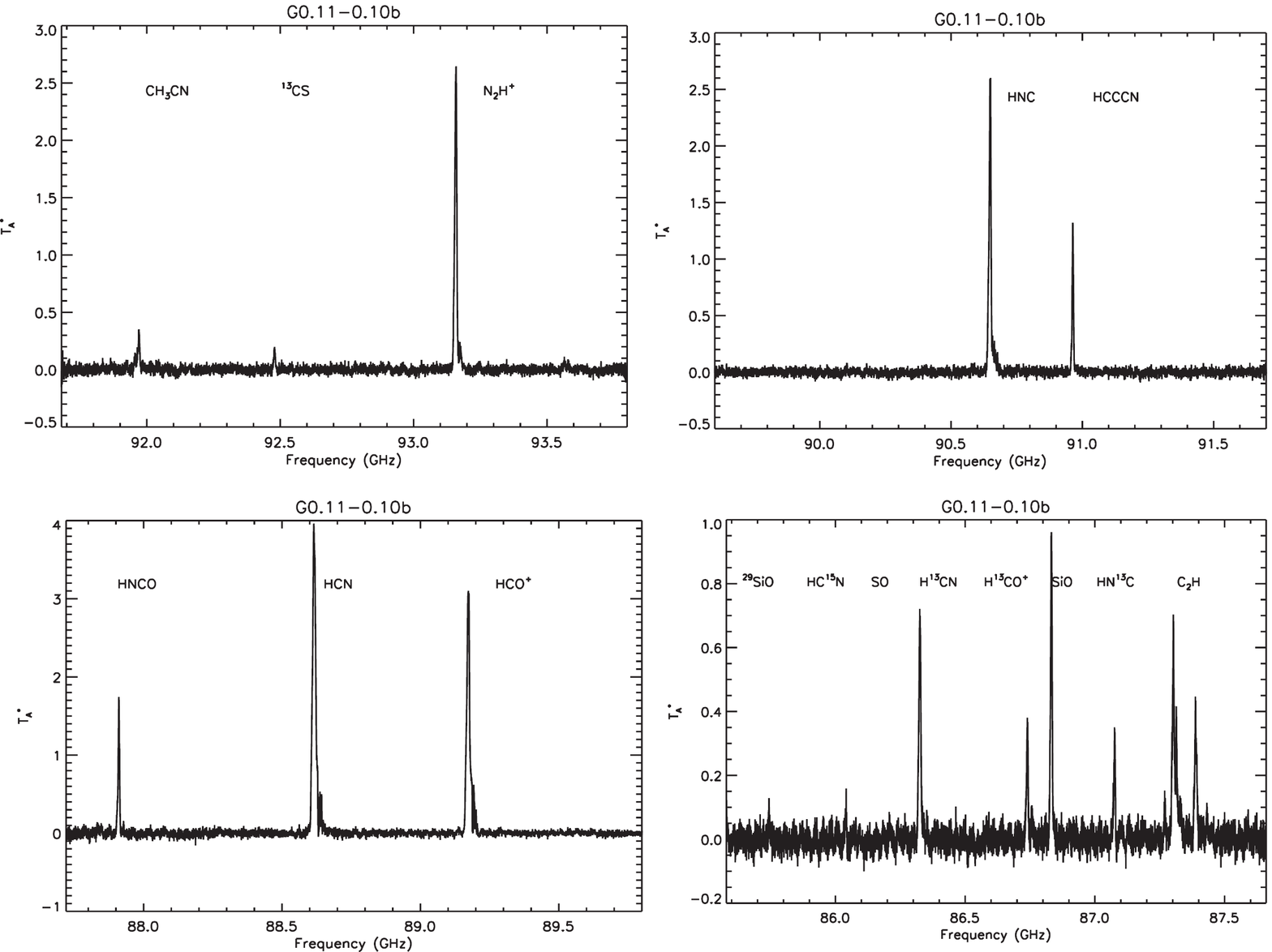}\\
\caption{
{\it (a - Top Left 4)} 
Deep pointed observations showing the spectra of 16 molecules 
toward position A  in G0.13--0.13 
{\it (b - Top Right 4)} Similar to {\it (a)} except for position B. 
{\it (c - Top Left 4)} Similar to   {\it (a)}  except for position C. 
{\it (d - Top Right 4)} Similar to   {\it (a)}  except for position D. 
}\end{figure}

% ******************************************
\setcounter{figure}{7} % reset figure counter to 1 so next figure is Figure 2 again.
% ******************************************
\begin{figure}[p] %  figure placement: here, top, bottom, or page
   \centering
\includegraphics[scale=0.3,angle=0]{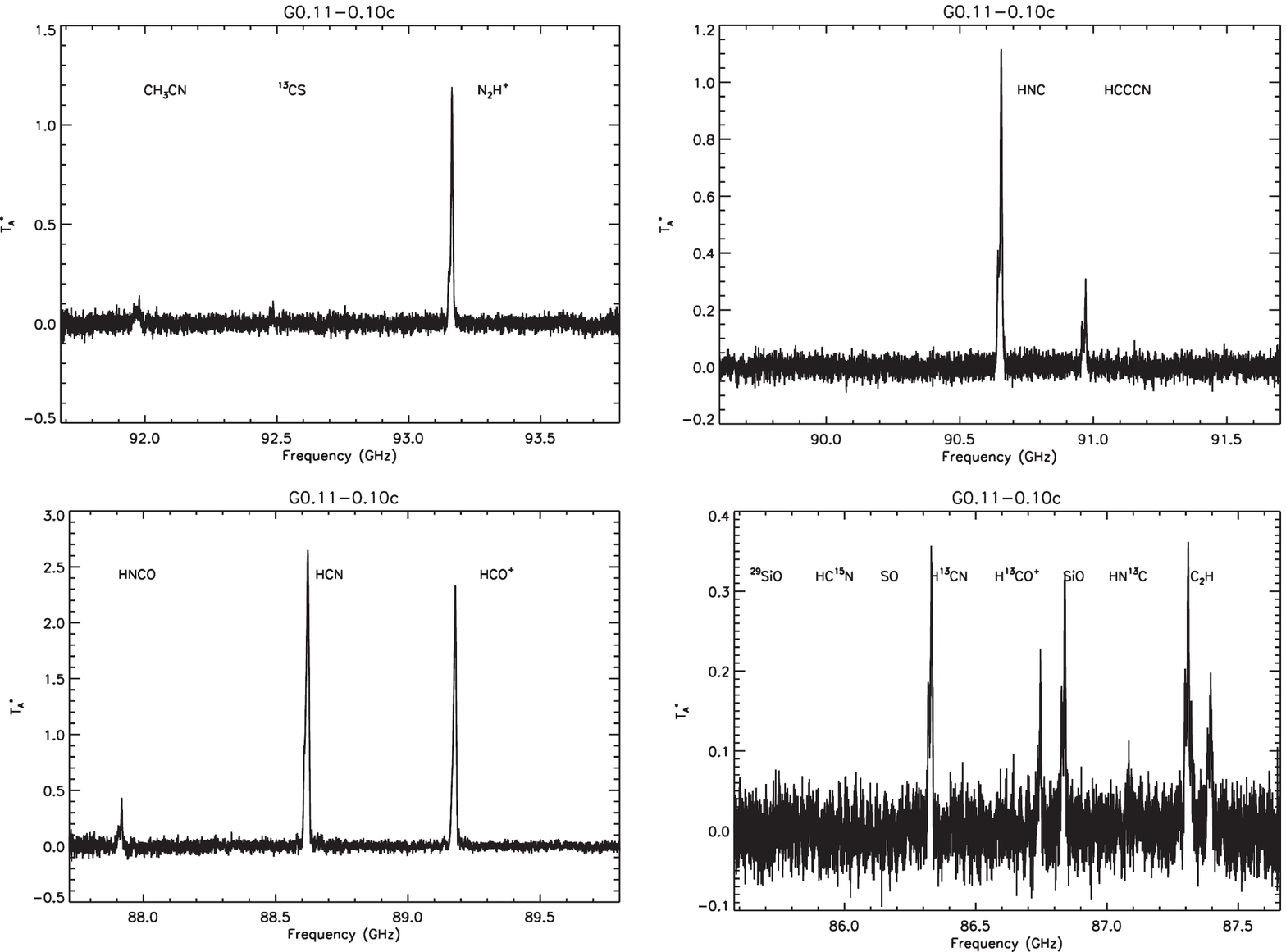}
\includegraphics[scale=0.3,angle=0]{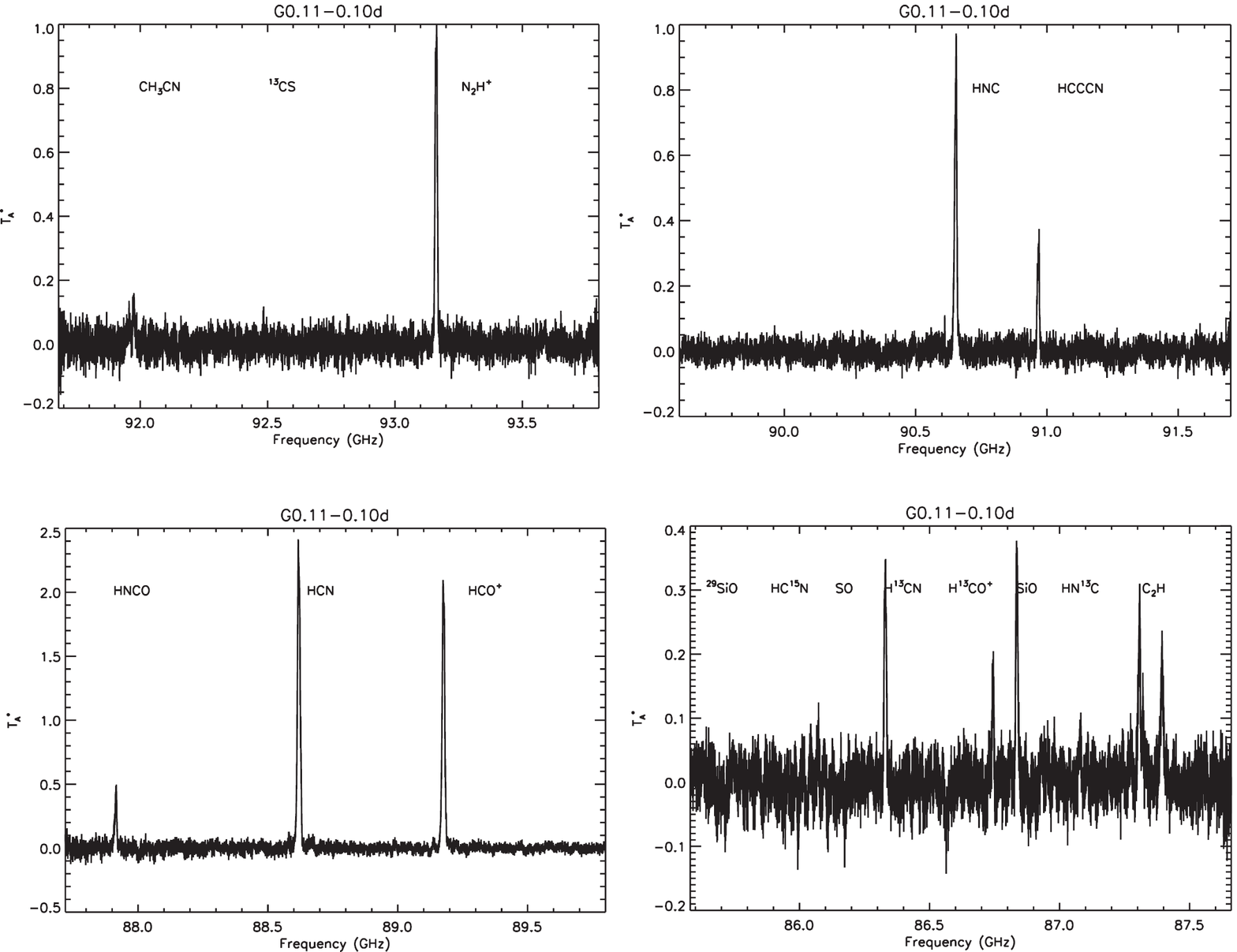}
   \caption{continued}
   \label{fig:example}
\end{figure}

\begin{figure}
\center
\includegraphics[scale=0.4,angle=0]{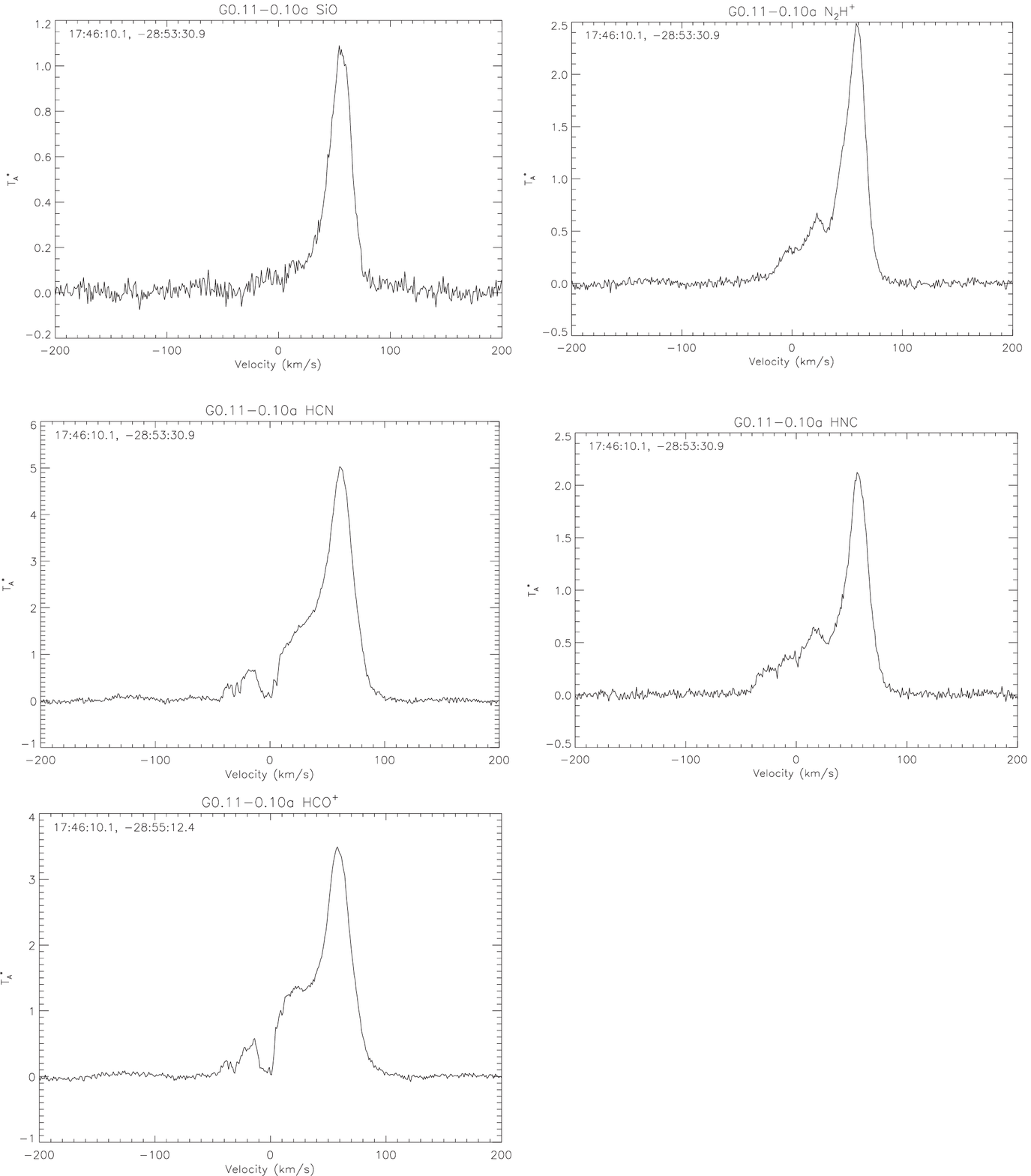}
\caption{Individual  spectrum 
 of five molecular species 
SiO, HCN, HNC, HCO$^+$ and N$_2$H$^+$ toward 
position A based on  Mopra observations. 
}
\end{figure}

\begin{figure}
\center
\includegraphics[scale=0.5,angle=0]{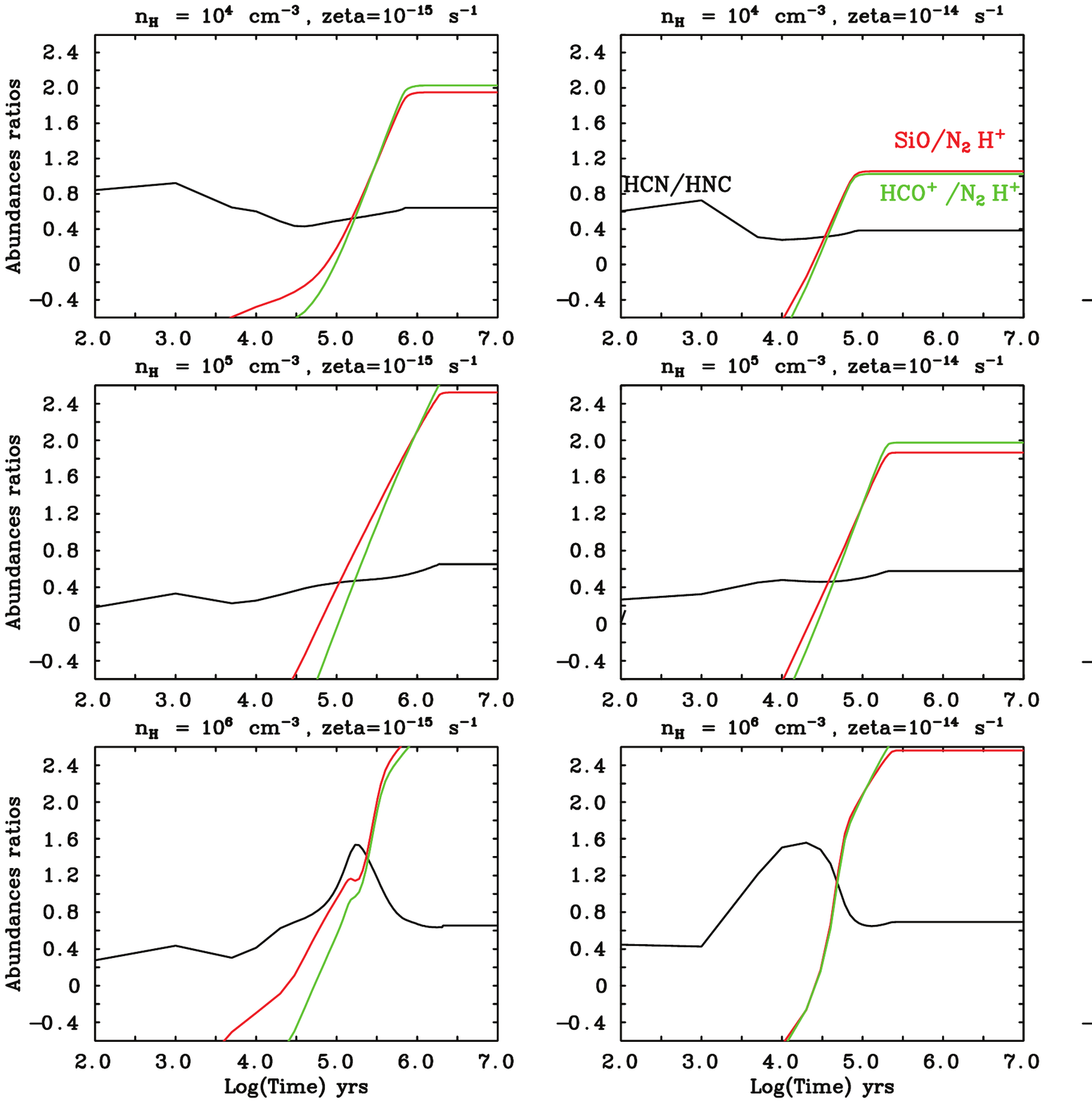}
\caption{Abundance ratios of 
HCN/HNC (black)  and HCO$^+$/N$_2$H$^+$ (green) 
and SiO/N$_2$H$^+$ (red) as a function of time in years for different 
values of cosmic ray ionization rate and molecular gas density. 
The chemical evolution of the G0.13--0.13 cloud 
uses the UCL\_CHEM  time-dependent gas-grain chemical model$^{31}$. 
Using the same color scheme in the model, 
the observed abundance ratios for all four positions A-D 
 are marked on the top right panel.  
%(Viti et al. 2004). 
}
\end{figure}

\vfill\eject
\begin{deluxetable}{c c c}
%\tabletypesize{\scriptsize}
\tablewidth{0pt}
\tablecaption{Observed Positions within G0.13--0.13}
\tablehead{
\colhead{Source}  & \colhead{$\ell$ } & \colhead{$b$} \\
\colhead{Name}  & \colhead{[$^{\circ}$]}   & \colhead{[$^{\circ}$]}
}\startdata
A    & 0.1007       & $-$0.0792 \\
B     & 0.0632       & $-$0.0709 \\
C     & 0.1132       & $-$0.1084 \\
D     & 0.1299       & $-$0.1292 \\
\enddata
\end{deluxetable}

\vfill\eject

\begin{deluxetable}{c c c c c c c c c c c}      % 11 columns
\rotate                                         % rotates table
\tabletypesize{\scriptsize}                     % gives the table smaller font
\tablewidth{0pt}                                % reduces the space between columns to ensure it fits
\tablecaption{Observed Peak Line Intensities, LSR  Velocities and their 1$\sigma$ Errors}
\tablehead{
\colhead{ Source }  &   
\colhead{ $v$(HCN) }  & \colhead{ $T^*_A$(HCN) }  
& \colhead{ $v$(HCO$^+$) }    &   \colhead{ $T^*_A$(HCO$^+$) }  & \colhead{ $v$(HNC) }    &   \colhead{ $T^*_A$(HNC) }    &   \colhead{ $v$(N$_2$H$^+$) } &   
 \colhead{ $T^*_A$(N$_2$H$^+$) } &   \colhead{ $v$(SiO) }    & 
  \colhead{ $T^*_A$(SiO) }  \\
\colhead{ Name }    &   \colhead{ km s$^{-1}$ } &   \colhead{ Peak (K) }    &   
\colhead{ km s$^{-1}$ } &   \colhead{ Peak (K) }    &   \colhead{ km s$^{-1}$ } &   \colhead{ Peak (K) }    &   \colhead{ km s$^{-1}$ } &   \colhead{ Peak (K) }   
 &   \colhead{ km s$^{-1}$ } &   \colhead{ Peak (K) }\\
}                           % end of tablehead
\startdata
A     & 60.65$\pm0.9$ & 5.02$\pm0.03$ & 57.91$\pm0.90$ & 3.49$\pm0.03$ & 54.92$\pm0.90$
 & 2.12$\pm0.02$  & 58.13$\pm0.90$ & 2.49$\pm0.02$ & 54.29$\pm0.90$ & 1.09$\pm0.03$  \\
B     & 57.88$\pm0.90$ & 3.96$\pm0.03$ & 51.53$\pm0.90$ & 3.10$\pm0.03$ & 50.42$\pm0.90$ & 2.59$\pm0.02$ 
& 48.54$\pm0.90$ & 2.64$\pm0.03$ & 49.60$\pm0.90$ & 0.96$\pm0.03$ \\
C     & 34.19$\pm0.90$ & 2.65$\pm0.03$ & 34.33$\pm0.90$ & 2.33$\pm0.03$ & 28.15$\pm0.90$ & 1.12$\pm0.03$ & 
32.08$\pm0.90$ & 1.19$\pm0.02$ & 26.36$\pm0.90$ & 0.32$\pm0.02$ \\
D     & 32.35$\pm0.90$ & 1.98$\pm0.03$ & 41.56$\pm0.90$ & 1.98$\pm0.03$ & 30.81$\pm0.90$ & 0.97$\pm0.03$
 & 33.80$\pm0.90$ & 1.00$\pm0.03$ & 42.1$\pm0.90$  & 0.38$\pm0.03$ \\
\enddata
\end{deluxetable}

\vfill\eject

\begin{deluxetable}{c c c c c c c }      % 7 columns
\rotate                                         % rotates table
\tabletypesize{\scriptsize}                     % gives the table smaller font
\tablewidth{0pt}                                % reduces the space between columns to ensure it fits
\tablecaption{Predicted Column Density from Observed  Intensity Ratios}
\tablehead{
 \colhead{ Source} 
& \colhead{[N(HCN)]/[N(HNC)]}  &  \colhead{ [N(HCO$^+$]/[N(N$_2H^+$)] }   & \colhead{ [N(SiO)]/N(N$_2H^+$)] } &
\colhead{ $T^*_A(HCN)/T^*_A(HNC)$ }  & \colhead{ $T^*_A(HCO^+)/T^*_A(N_2H^+$) } & \colhead{ $T^*_A(SiO)/T^*_A(N_2H^+$) }\\
}
\startdata
A     & 0.33$\pm0.01$ & 0.13$\pm0.01$ & -0.10$\pm0.01$  & 2.37$\pm0.03$ & 1.40$\pm0.02$ & 0.44$\pm0.01$ \\
B     & 0.01$\pm0.001$ & 0.005$\pm0.001$ & -0.20$\pm0.02$ & 1.53$\pm0.02$ & 1.17$\pm0.02$ & 0.36$\pm0.01$  \\
C     & 0.35$\pm0.005$ & 0.19$\pm0.01$ & -0.41$\pm0.03$  & 2.37$\pm0.05$ & 1.96$\pm0.04$ & 0.27$\pm0.02$  \\ 
D     & 0.26$\pm0.01$ & 0.16$\pm0.01$ & -0.24$\pm0.03$  & 2.04$\pm0.06$ & 1.98$\pm0.07$ & 0.38$\pm0.03$  \\
\enddata
\end{deluxetable}

\vfil\eject

\end{document}